\documentclass[prd,aps, tightenlines, preprintnumbers, showpacs,nofootinbib,superscriptaddress,notitlepage]{revtex4-1}

\usepackage{graphicx}
\usepackage{amsmath,amsthm,amssymb}

\newcommand{\non}{\nonumber\\}

\begin{document}
\title{Forward Hadron Productions in Proton-Proton Collisions in Small-$x$ Formalism}

\author{Kazuhiro Watanabe}
\author{Bo-Wen Xiao}
\affiliation{Key Laboratory of Quark and Lepton Physics (MOE) and Institute
of Particle Physics, Central China Normal University, Wuhan 430079, China}

\begin{abstract}
Employing the so-called hybrid formalism, we calculate the cross section of inclusive hadron production in proton-proton collisions at forward rapidity in small-$x$ formalism at one-loop order. For the case of hadron production at forward rapidity, we can uses collinear parton distributions for projectile proton and $k_\perp$ dependent gluon distribution for target proton. We show that collinear divergences associated with initial and final state parton radiations are renormalized into parton distributions and fragmentation functions in terms of the Dokshitzer-Gribov-Lipatov-Altarelli-Parisi evolution equation, respectively. Furthermore, rapidity divergence can be absorbed into the wave function of target proton which gives rise to the well-known Balitsky-Fadin-Kuraev-Lipatov equation. These divergences are completely separated from the short distance partonic hard parts, which is now finite at the next-to-leading order accuracy. The result presented in this paper can be reckoned as a baseline calculation without any non-linear QCD effects in small-$x$ formalism. As a consistency check, we compare our results with the previous calculation for non-linear proton-nucleus collisions in the small-$x$ formalism and find complete agreement in the dilute and large $N_c$ limit. In phenomenology, the direct comparison of the above two separate calculations can reveal the role and strength of the non-linear dynamics in high energy QCD, and thus help us reliably study the onset of gluon saturation when genuine non-linear interactions become important.  
\end{abstract}


\maketitle

\section{Introduction}

In high energy scatterings, gluon bremsstrahlung radiation plays a crucial role in describing the rapid rise of gluon density inside hadron and nucleus. The soft gluon radiation is ingeniously encoded in the Balitsky-Fadin-Kuraev-Lipatov (BFKL) evolution equation~\cite{Balitsky:1978ic,Lipatov:1976zz,Kuraev:1976ge,Fadin:1975cb} which resums small-$x$ logarithms. As a result, the energy evolution of off-shell unintegrated parton distribution function in dilute limit is governed by the BFKL equation in contrast to the well known Dokshitzer-Gribov-Lipatov-Altarelli-Parisi (DGLAP) evolution for collinear parton densities~\cite{Gribov:1972ri,Altarelli:1977zs,Dokshitzer:1977sg}. In phenomenology, to describe particle productions in hadronic collisions in the small $x$ and low transverse momentum transfer region, we need to take into account the off-shellness of incident partons~\cite{Collins:1991ty,Catani:1990eg}, and switch from collinear factorization framework to $k_t$-factorization framework (e.g., Ref.~\cite{Kharzeev:2003wz,Kharzeev:2004yx,Blaizot:2004wu}). 

Study of inclusive particle productions in high energy hadron-nucleus/nucleus-nucleus collisions has attracted a lot of attention in the past few years, since this process can help to reveal nonlinear gluon dynamics on top of the linear BFKL evolution when gluon density inside large nucleus is sufficiently high. As collision energy increases, the gluon occupation number in the low-$x$ region inside nucleus becomes large due to gluon bremsstrahlung radiation. The subsequent balance between the gluon bremsstrahlung and recombination leads to the phenomenon of gluon saturation~\cite{Gribov:1984tu,Mueller:1985wy, Mueller:2001fv} or color-glass-condensate~\cite{Weigert:2005us,Gelis:2010nm,Kovchegov:2012mbw}. In principle, the so-called saturation scale $Q_s$ separates the nonlinear dense regime $Q\leq Q_s$ from dilute regime $Q\gg Q_s$ with $Q$ being the typical scale of the external probe. Since $Q_s$ is enhanced by a factor $A^{1/3}$ for heavy nuclei with $A$ being the number of nucleon, current proton-nucleus (p$A$) collision and a future Electron-Ion-Collider are expected to provide unique opportunities to investigate the gluon saturation phenomenon. In particular, the measurement of the nuclear modification factor $R_{{\rm p}A}\equiv\frac{1}{A}\frac{d\sigma_{{\rm p}A}/d^2p_{\perp}dy}{d\sigma_{{\rm pp}}/d^2p_{\perp}dy}$ at RHIC and the LHC~\cite{Arsene:2004ux,Adams:2003qm,ALICE:2012mj} reflects the relatively different strength of the saturation phenomenon in pp and p$A$ collisions and it has been described by various theoretical calculations~\cite{Kharzeev:2004yx,Albacete:2010bs,Albacete:2012xq,Lappi:2013zma,JalilianMarian:2011dt,Albacete:2014fwa} in small-$x$ physics.

Furthermore, inclusive single hadron production in p$A$ collisions (${\rm p}+A\rightarrow h+X$) has been studied in the small-$x$ formalism based on the Mueller's dipole model~\cite{Mueller:1993rr,Mueller:1994gb}. In particular, at forward rapidity, the longitudinal momentum fraction of target nucleus carried by incoming parton becomes small and the saturation scale for heavy nuclei is much larger than typical hadronic scale $\Lambda_\text{QCD}$, whereas the saturation scale of the proton projectile is expected to be much smaller. Therefore, as long as we focus on the forward hadron production, the so-called hybrid treatment, in which the collinear parton coming from projectile proton scatters off dense gluons inside target nucleus with finite transverse momentum transfer, is considered to be a reasonable and simple approach~\cite{Dumitru:2002qt,Dumitru:2005kb}. In the hybrid formalism, hadron productions in p$A$ collisions are given by the convolution of short distance hard parts, collinear parton distribution functions (PDFs) for projectile proton, fragmentation functions (FFs) for produced hadron, and the wave function of target nucleus constructed from multi-point Wilson line correlators. In general, multi-point Wilson line correlators obey the so-called JIMWLK evolution equation~\cite{Jalilian-Marian:1997jx+X}. In phenomenological studies, large-$N_c$ and mean field approximations are usually utilized to reduce multi-point Wilson line correlators to products of dipole correlators which resum the small-$x$ logarithm $\alpha_s\ln1/x$ by means of the Balitsky-Kovchegov (BK) equation~\cite{Balitsky:1995ub,Kovchegov:1996ty}.

In addition, inclusive hadron production in p$A$ collisions beyond leading order (LO) has been firstly studied in Refs.~\cite{Albacete:2010bs,JalilianMarian:2011dt,Altinoluk:2011qy,Fujii:2011fh} where the running coupling BK equation is employed together with the LO framework. These studies included a subset of next-to-leading order (NLO) corrections. The pursuit of complete NLO calculations began several years ago~\cite{Chirilli:2011km,Chirilli:2012jd,Stasto:2013cha,Altinoluk:2014eka,Watanabe:2015tja,Albacete:2014fwa,Ducloue:2016shw}. In Refs.~\cite{Chirilli:2011km,Chirilli:2012jd}, the systematic NLO calculations is carried out for the first time at the high energy limit. It is shown that the collinear divergences and the rapidity divergences, which arise from one-loop diagrams at NLO, are clearly separable from the partonic hard scattering part in the small-$x$ formalism for inclusive hadron production in p$A$ collisions. This allows us to establish an effective factorization in the hybrid formalism at the one-loop order. The small-$x$ factorization formula for inclusive hadron production at forward rapidity $y$ with transverse momentum $p_{\perp}$ in p$A$ collisions can be cast into~\cite{Chirilli:2012jd}
\begin{align}
\frac{d\sigma^{{\rm p}+A\rightarrow h+X}}{d^2p_{\perp}dy}
=\sum_{i=q,g}\int \frac{dz}{z^2}x_pf_i\left(x_p,\mu\right){\cal F}^{F,A}_{x_g}(k_\perp)D_{h/i}(z,\mu)+\sum_{i,j=q,g}\frac{\alpha_s}{2\pi}\int\frac{dz}{z^2}\frac{dx}{x}\xi xf_i\left(x,\mu\right)S_{ij}D_{h/j}(z,\mu)
\label{eq:factorization}
\end{align}
where $x_p=k_\perp e^y/\sqrt{s}$ is longitudinal momentum fraction of projectile proton carried by incoming parton, $k_\perp$ is transverse momentum of gluon coming from target nucleus with longitudinal momentum $x_g=k_\perp e^{-y}/\sqrt{s}$, and $\xi=x_p/x$. In addition, $z$ is momentum fraction of parton carried by produced hadron in final state. $f_i(x)$ is the collinear PDF of projectile proton and $D_{h/i}(z)$ is the FF from parton $i$ to hadron $h$. ${\cal F}^{F,A}_{x_g}$ is Fourier transform of the dipole amplitude in the fundamental or adjoint representation which is determined by the partonic scattering at LO. This dipole amplitude encodes all the important information on the scattering between incoming partons and target harons. $S_{ij}$ represent the partonic scattering processes at NLO convoluted with the dipole amplitude. The factorization scale $\mu$ dependence of the PDFs and the FFs is derived from the DGLAP equation due to quantum evolution, while the rapidity or energy dependence of the dipole amplitude is described by the BK equation.

In Ref.~\cite{Watanabe:2015tja}, the improved numerical calculation for inclusive hadron production in p$A$ collisions at NLO is performed and it yields the inclusive hadron spectra which are in excellent agreement with both RHIC and the LHC data for various kinematical regions. Nevertheless, it is still hard to tell whether this agreement is due to the non-linear gluon saturation phenomenon or the linear BFKL dynamics, since we do not know for sure whether the saturation effect is indispensable to describe the current data in pp/p$A$ collisions or not. In order to disentangle these two effects, it is very useful to conduct a comparative NLO calculation for single hadron production in pp collisions with only linear BFKL dynamics, which would help us clearly visualize the different behaviour of linear and non-linear dynamics in the hadron $p_\perp$ spectra and eventually lead us to better understanding of the onset of gluon saturation phenomenologically. The objective of this paper is to calculate the NLO forward inclusive hadron production in pp collisions in the small-$x$ formalism in the dilute regime. As a cross check, we find that the results that we obtain for pp collisions agree with the previous calculation for p$A$ collisions (assuming large $N_c$ approximation)~\cite{Chirilli:2012jd, Watanabe:2015tja} in the large $N_c$ and dilute limit. 

Through this paper, we use the light cone perturbation theory~\cite{Lepage:1980fj} together with the light-cone gauge $A^+=0$ and light-cone coordinates $x^{\pm}=(x^0\pm x^3)/{\sqrt 2}$ and $x^{\mu}=(x^+,x^-,x_\perp)$ with the metric $g_{+-}=g_{-+}=1$, $g_{11}=g_{22}=-1$. 

The rest of this paper is organized as follows: In Sec.~II, we firstly consider inclusive hadron production in pp collisions at LO within the hybrid formalism. Next, we calculate four NLO channels in Sec.~III. We show that the rapidity divergences and the collinear divergences can be completely factorized from the partonic hard scattering part like Eq.~(\ref{eq:factorization}). We see that the factorization scale dependence of the PDF and the FF is controlled by the DGLAP equation and the energy dependence of the dipole amplitude is described by the BFKL equation. The main result of this paper is given in Eq.~(\ref{eq:master-expression}). Finally, in Sec.~IV, we summarize our calculation and discuss its future development.


\section{The Leading Order}

\begin{figure}
\centering
\includegraphics[width=12cm,angle=0]{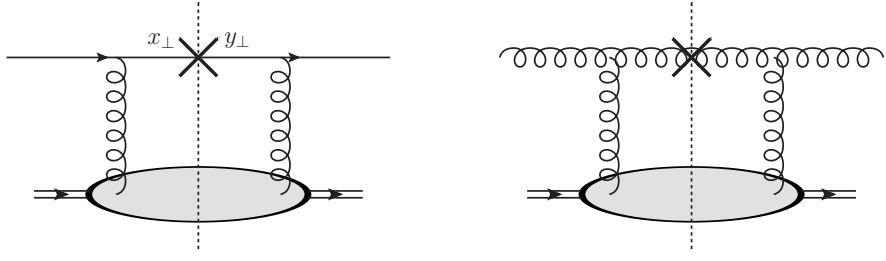}
\caption{Lowest order diagrams for $q+{\rm p}\rightarrow q+X$ (left) and $g+{\rm p}\rightarrow g+X$ (right) with one gluon exchange in the $t$-channel. 
 }
\label{fig:LO-diagram}
\end{figure}

In this section, let us first consider the LO forward single hadron production in pp collisions. For pp collision, the center of mass energy is $s=(P+P^\prime)^2\simeq P^+P^{\prime-}$ with $P$ ($P^\prime$) being momentum of projectile (target) proton.  Here the projectile (target) proton beam is supposed to have large light cone plus (minus) component and then $P^-=P^{\prime+}\simeq 0$ and $P_\perp=P^{\prime}_\perp\simeq 0$. We begin with a basic process of quark scattering off proton ($q+{\rm p}\rightarrow q+X$). 
In the small-$x$ formalism, the differential cross section for producing a quark with momentum $k=(k^+,k_\perp)$ is given by
\begin{align}
\frac{d\sigma^{q+{\rm p}\rightarrow q+X}_{\rm LO}}{d^3k}=\delta(p^+-k^+)\int\frac{d^2x_\perp d^2y_\perp}{(2\pi)^2}e^{-ik_\perp\cdot(x_\perp-y_\perp)}S_{x_g}(x_\perp,y_\perp)
\end{align}
where we have averaged over spin and color of the incoming quark and summed all of the quantum number of the final state, $p^+$ is the light cone momentum of the incoming quark and $k^+$ is the momentum of the observed quark. The transverse momentum of incident parton is assumed to be $p_\perp=0_\perp$ for the sake of simplicity. The delta function is due to the momentum conservation. $S_{x_g}$ is the color singlet dipole scattering amplitude in the fundamental representation
\begin{align}
S_{x_g}(x_\perp,y_\perp)=\frac{1}{N_c}\langle {\rm Tr}\left[U(x_\perp)U^\dagger(y_\perp)\right]\rangle_{x_g}
\end{align}
with $r_\perp=x_\perp-y_\perp$ being a transverse size of the dipole (see FIG.~\ref{fig:LO-diagram}). 
$U(x_\perp)$ is the fundamental Wilson line at transverse coordinate $x_\perp$ which resums multiple scatterings between the quark and the gluon of target hadron in the eikonal approximation. In dilute regime, one expects 
\begin{align}
U(x_\perp)={\cal P}\exp\left[ig\int dx^+ t^aA_a^-(x^+,x_\perp)\right] \approx 1+ig\int dx^+ t^aA_a^-(x^+,x_\perp),
\end{align}
where $A(x^+,x_\perp)$ is the effective background gauge field of the target proton at $x=(x^+,x_\perp)$ and $t^a$ is the generator of SU(3) group. $x_g=k^-/P^{\prime-}$ is the longitudinal momentum fraction of the target proton carried by the small-$x$ gluon attached to the quark. $\langle\rangle_{x_g}$ represents the average over all of the color configurations of the gluon inside the target proton. To obtain the cross section for ${\rm p}+{\rm p}\rightarrow q+X$, we can couple the above results with collinear quark PDF $q_f$ with flavor $f$ and write
\begin{align}
\frac{d\sigma^{{\rm p}+{\rm p}\rightarrow q+X}_{\rm LO}}{d^3k}=\sum_f\int dx q_f(x)\frac{d\sigma^{q+{\rm p}\rightarrow q+X}_{\rm LO}}{d^3k}
\end{align}
with $x_p=p^+/P^+$ being the longitudinal momentum fraction of the projectile proton carried by the incoming parton.
Furthermore, together with the collinear FF which converts $q$ into hadron $h$ in the final state, the differential cross section of single hadron production with $p_{h\perp}$ at rapidity $y$ is given by
\begin{align}
\frac{d\sigma_{\rm LO}^{{\rm p}+{\rm p}\rightarrow h/q+X}}{d^2p_{h\perp} dy}=\sum_f\int_\tau^1\frac{dz}{z^2}D_{h/q}(z)x_pq_f\left(x_p\right){\cal F}_{x_g}(k_{\perp})
\label{eq:lo-q-momentum}
\end{align}
with $zk_\perp=p_{h\perp}$, $\tau=zx_p$, $x_p=k_\perp e^{+y}/\sqrt{s}$, and $x_g=k_\perp e^{-y}/\sqrt{s}$. 
${\cal F}_{x_g}$ is the Fourier transform of the dipole amplitude in the fundamental representation defined as
\begin{align}
{\cal F}_{x_g}(k_{\perp})
\equiv\int\frac{d^2x_\perp d^2y_\perp}{(2\pi)^2}e^{-ik_\perp\cdot(x_\perp-y_\perp)}(-T_{x_g}(x_\perp,y_\perp))=S_\perp F_{x_g}(k_\perp),
\label{fundamental-amplitude}
\end{align}
where $F_{x_g}(k_{\perp})=\int\frac{d^2r_\perp}{(2\pi)^2}e^{-ik_\perp\cdot r_\perp}(-T_{x_g}(r_\perp))$ with $T_{x_g}=1-S_{x_g}$ being the forward scattering amplitude. We can write the cross section which is proportional to the transverse area of the target proton $S_\perp$ if the impact parameter dependence is neglected. We note that we have dropped the elastic part which is proportional to $\delta^{(2)}(k_\perp)$, since we are only interested in the inelastic production with finite $k_\perp$. 

Similarly, for the gluon channel at LO, one finds
\begin{align}
\frac{d\sigma_{\rm LO}^{{\rm p}+{\rm p}\rightarrow h/g+X}}{d^2p_{h\perp} dy}=\int_\tau^1\frac{dz}{z^2}D_{h/g}(z)x_pG\left(x_p\right)\widetilde{{\cal F}}_{x_g}(k_{\perp})
\label{eq:lo-g-momentum}
\end{align}
where $G$ and $D_{h/g}$ are the collinear gluon PDF and FF, respectively. The Fourier transform of the forward scattering amplitude in the adjoint representation is
\begin{align}
\widetilde{{\cal F}}_{x_g}(k_{\perp})
=\int\frac{d^2x_\perp d^2y_\perp}{(2\pi)^2}e^{-ik_\perp\cdot(x_\perp-y_\perp)}(-\widetilde{T}_{x_g}(x_\perp,y_\perp)).
\end{align}
with
\begin{align}
\widetilde{S}_{x_g}(x_\perp, y_\perp)=1-\widetilde{T}_{x_g}(x_\perp, y_\perp)=\frac{1}{N_c^2-1}\langle {\rm Tr}\left[W(x_\perp)W^\dagger(y_\perp)\right]\rangle_{x_g}
\end{align}
where $W(x_\perp)$ is the adjoint Wilson line defined at $x_\perp$. Using the identity
\begin{align}
W^{ab}(x_\perp)=2{\rm Tr}\left[t^aU(x_\perp)t^bU^\dagger(x_\perp)\right],
\end{align}
we can approximate the adjoint dipole amplitude in the dilute regime of target proton as 
\begin{align}
\widetilde{S}_{x_g}(x_\perp, y_\perp)&=\frac{1}{N_c^2-1}\left[\langle{\rm Tr}\left[U(x_\perp)U^{\dagger}(y_\perp)\right]{\rm Tr}\left[U(y_\perp)U^{\dagger}(x_\perp)\right]\rangle_{x_g}-1\right]\non
&\approx1-\frac{N_c}{C_F}T_{x_g}(x_\perp,y_\perp).
\end{align}
Hereafter, we will use $\widetilde{T}_{x_g}\equiv N_cT_{x_g}/C_F$ in the following caculation.


\section{The Next to Leading Order}

Let us now consider the corresponding NLO calculation which can be put into four different channels. Essentially, the calculations which we perform below are similar to the calculations in Ref.~\cite{Chirilli:2012jd}. However, we would like to point out that there are some subtleties in order to obtain the differential cross section for inclusive hadron production in pp collisions at NLO at finite-$N_c$. In fact, it seems that the following calculation of this process in pp collisions in the dilute regime is slightly harder than the calculation which leads to the non-linear results obtained in p$A$ collisions, since the large $N_c$ approximation was employed in the latter case.

\subsection{The $q\rightarrow q$ channel}

\begin{figure}
\centering
\includegraphics[width=16cm,angle=0]{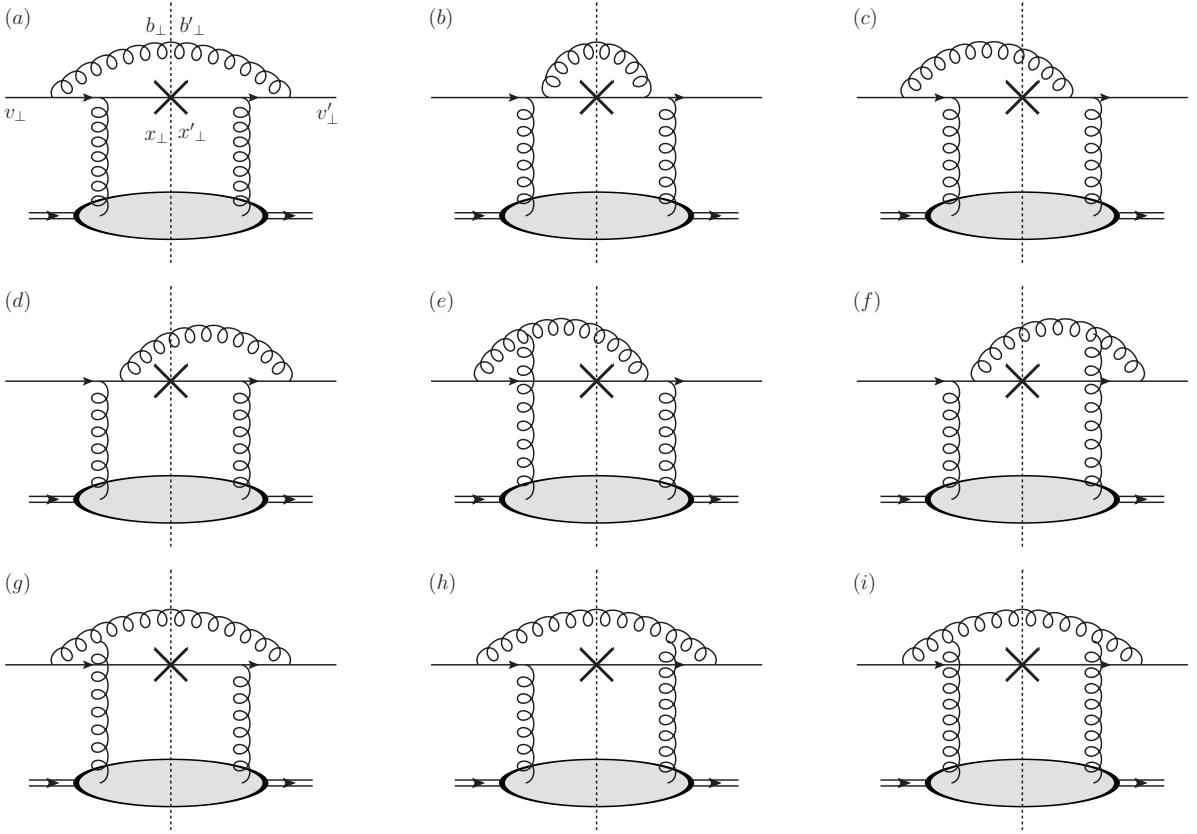}
\caption{Real diagrams at NLO for the $q\rightarrow q$ channel. Notations are the same as FIG.~\ref{fig:LO-diagram}.
 }
\label{fig:NLO-qq}
\end{figure}

For $q+{\rm p}\rightarrow q+g+X$ channel as shown in FIG.~\ref{fig:NLO-qq}, the differential cross section for producing a quark with momentum $k$ and a gluon with $l$ is given by
\begin{align}
\frac{d\sigma^{q+{\rm p}\rightarrow q+g+X}}{d^3kd^3l}
=&\alpha_s\delta(p^+-k^+-l^+)\int\frac{d^2x_\perp d^2x_\perp^\prime d^2b_\perp d^2b_\perp^\prime}{(2\pi)^8}
e^{-ik_{\perp}\cdot(x_\perp-x_\perp^\prime)}e^{-il_{\perp}\cdot(b_\perp-b_\perp^\prime)}\sum_{\alpha\beta\lambda}\psi_{qg\alpha\beta}^{\lambda\ast}(u_\perp^\prime)\psi_{qg\alpha\beta}^{\lambda}(u_\perp)\non
&\times
\Bigg[C_FS_{x_g}(x_\perp,x_\perp^\prime)+C_FS_{x_g}(v_\perp,v_\perp^\prime)+\frac{1}{2N_c}S_{x_g}(x_\perp,v_\perp^\prime)+\frac{1}{2N_c}S_{x_g}(v_\perp,x_\perp^\prime)\non
&-\frac{N_c}{2}S_{x_g}(v_\perp,b_\perp)-\frac{N_c}{2}S_{x_g}(b_\perp,v_\perp^\prime)-\frac{N_c}{2}S_{x_g}(x_\perp,b_\perp)-\frac{N_c}{2}S_{x_g}(b_\perp,x_\perp^\prime)+N_cS_{x_g}(b_\perp,b_\perp^\prime)\Bigg]
\end{align}
where $u_\perp=b_\perp-x_\perp$ and $u_\perp^\prime=b_\perp^\prime-x_\perp^\prime$ are the transverse distance between the produced gluon and the quark in the amplitude and the complex conjugate amplitude, respectively. $v_\perp=\xi x_\perp+(1-\xi)b_\perp$ and $v_\perp^\prime=\xi x_\perp^\prime+(1-\xi)b_\perp^\prime$ are the transverse coordinate of the incoming quark in the amplitude and the complex conjugate amplitude, respectively. $\xi=k^+/p^+$ is the longitudinal momentum fraction of the incoming quark carried by the produced quark in the final state. $\psi_{qg\alpha\beta}^{\lambda}$ is the light cone wave function which describes the quark-gluon splitting amplitude~\cite{Marquet:2007vb}
\begin{align}
\psi_{qg\alpha\beta}^\lambda(u_\perp)=2\pi i\sqrt{\frac{2}{(1-\xi)p^+}}
\left\{
\begin{array}{c}
\frac{u_\perp\cdot \varepsilon^{(1)}_\perp}{u_\perp^2}(\delta_{\alpha-\beta-}+\xi\delta_{\alpha+}\delta_{\beta+})~~~(\lambda=1)\\
\frac{u_\perp\cdot \varepsilon^{(2)}_\perp}{u_\perp^2}(\delta_{\alpha+\beta+}+\xi\delta_{\alpha-}\delta_{\beta-})~~~(\lambda=2),
\end{array}
\right.
\end{align}
where $\alpha$ and $\beta$ are a spin of the incoming quark and the outgoing quark, respectively. $\lambda$ is a polarization of the radiated gluon. The polarization vector is defined as ${\varepsilon_\perp^{(\lambda)}}=-\frac{1}{\sqrt2}(\sigma, i)$ with $\sigma=+1$ for $\lambda=1$ and $\sigma=-1$ for $\lambda=2$, respectively.
Summing over $\alpha$, $\beta$, and $\lambda$, one finds 
\begin{align}
\sum_{\alpha\beta\lambda}\psi_{qg\alpha\beta}^{\lambda\ast}(u_\perp^\prime)\psi_{qg\alpha\beta}^{\lambda}(u_\perp)
=\frac{2(2\pi)^2}{p^+}\frac{1+\xi^2}{1-\xi}\frac{u_\perp^\prime\cdot u_\perp}{u_\perp^{\prime2}u_{\perp}^2}.
\label{eq:splitting-kernel-qq}
\end{align}
Summing over all the real diagrams, the total real contribution for inclusive hadron production reads
\begin{align}
\frac{\alpha_s}{2\pi^2}&\int^1_\tau\frac{dz}{z^2}D_{h/q}(z)\int d^2l_{\perp}\int^1_{\frac{\tau}{z}}d\xi \frac{1+\xi^2}{1-\xi}\frac{x_p}{\xi}q_f\left(\frac{x_p}{\xi}\right)\int\frac{d^2x_\perp d^2x_\perp^\prime d^2b_\perp d^2b_\perp^\prime}{(2\pi)^4}
e^{-ik_{\perp}\cdot(x_\perp-x_\perp^\prime)}e^{-il_{\perp}\cdot(b_\perp-b_\perp^\prime)}\frac{u_\perp^\prime\cdot u_\perp}{u_\perp^{\prime2}u_\perp^2}\non
\times
\Bigg[&-C_FT_{x_g}(x_\perp,x_\perp^\prime)-C_FT_{x_g}(v_\perp,v_\perp^\prime)+\frac{N_c}{2}T_{x_g}(v_\perp,b_\perp)+\frac{N_c}{2}T_{x_g}(b_\perp,v_\perp^\prime)+\frac{N_c}{2}T_{x_g}(x_\perp,b_\perp)+\frac{N_c}{2}T_{x_g}(b_\perp,x_\perp^\prime)\non
&-\frac{1}{2N_c}T_{x_g}(x_\perp,v_\perp^\prime)-\frac{1}{2N_c}T_{x_g}(v_\perp,x_\perp^\prime)-N_cT_{x_g}(b_\perp,b_\perp^\prime)\Bigg],
\label{eq:qq-real2}
\end{align}
where we have not yet integrated over $l_{\perp}$ since the upper limit of $\xi$ actually depends on $l_\perp$, as we will consider below.
We note here that FIG.~\ref{fig:NLO-qq} $(i)$ eventually does not contribute the inelastic single hadron production thank to the unitarity constraint~\cite{Mueller:2001fv}.

\begin{figure}
\centering
\includegraphics[width=18cm,angle=0]{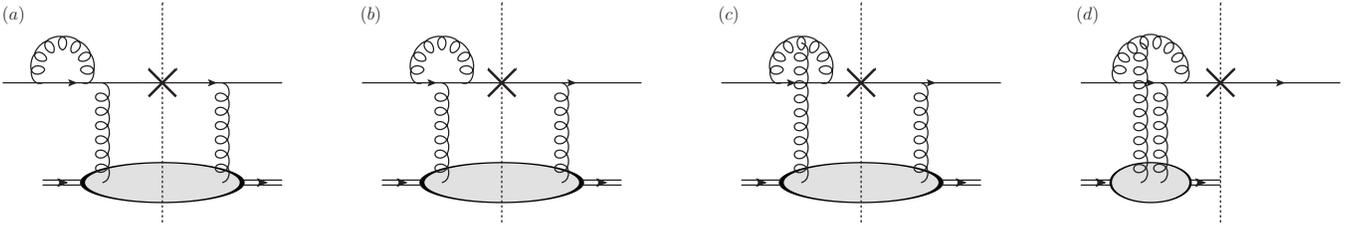}
\caption{Virtual diagrams at NLO for the $q\rightarrow q$ channel. Mirror contributions are omitted.
 }
\label{fig:NLO-qq-virtual}
\end{figure}

In the meantime, the virtual diagrams shown in FIG.~\ref{fig:NLO-qq-virtual} yield the following contribution
\begin{align}
-2\frac{\alpha_s}{2\pi^2}&\int^1_\tau\frac{dz}{z^2}D_{h/q}(z)x_pq_f(x_p)\int_0^1d\xi\frac{1+\xi^2}{1-\xi}\int\frac{d^2v_\perp d^2v_\perp^\prime d^2u_\perp}{(2\pi^)}e^{-ik_{\perp}\cdot(v_\perp-v_\perp^\prime)}\frac{1}{u_\perp^2}\non
&\times\Bigg[-C_FT_{x_g}(v_\perp,v_\perp^\prime)-\frac{1}{2N_c}T_{x_g}(x_\perp,v_\perp^\prime)+\frac{N_c}{2}T_{x_g}(b_\perp,v_\perp^\prime)+\frac{N_c}{2}T_{x_g}(b_\perp,x_\perp)\Bigg]
\label{eq:qq-virtual}
\end{align}
where we multiplied a factor of 2 explicitly to reflect the mirror diagrams of FIG.~\ref{fig:NLO-qq-virtual}. We would like to emphasize that an unfamiliar contribution from FIG.~\ref{fig:NLO-qq-virtual} $(d)$ plays an important role at $\xi\rightarrow 1$ due to the unitarity relation~\cite{Mueller:2001fv}, albeit it does not contribute to the inelastic hadron production as we explain below. 

It is manifest that the above results have the so-called rapidity singularities at $\xi=1$ in which the radiated gluon becomes soft. To deal with the rapidity singularities we adopt the plus function which is defined as follows
\begin{align}
\int_\tau^1d\xi\frac{1+\xi^2}{1-\xi}F(\xi)=\int_\tau^1d\xi\frac{1+\xi^2}{(1-\xi)_+}F(\xi)+\int_0^1d\xi\frac{2}{1-\xi}F(1)
\label{eq:plus-function}
\end{align}
for arbitrary function $F$. Making use of Eq.~(\ref{eq:plus-function}) and the identity 
\begin{align}
\frac{x_\perp}{x_\perp^2}=\int\frac{d^2k_\perp}{2\pi i}\frac{k_\perp}{k_\perp^2}e^{ik_\perp\cdot x_\perp},
\end{align}
the nonsingular part in the momentum space can be cast into
\begin{align}
&\frac{\alpha_s}{2\pi^2}\int_\tau^1\frac{dz}{z^2}D_{h/q}(z)\Bigg\{\int^1_{\frac{\tau}{z}} d\xi\frac{1+\xi^2}{(1-\xi)_+}\frac{x_p}{\xi}q_f\left(\frac{x_p}{\xi}\right)\int d^2k_{g\perp}{\cal F}_{x_g}(k_{g\perp})\Bigg[C_F\frac{1}{(k_{\perp}-k_{g\perp})^2}+C_F\frac{1}{(k_{\perp}-\xi k_{g\perp})^2}\non
&-N_c\frac{k_\perp\cdot(k_\perp-k_{g\perp})}{k_\perp^2(k_\perp-k_{g\perp})^2}-N_c\frac{k_\perp\cdot(k_\perp-\xi k_{g\perp})}{k_\perp^2(k_\perp-\xi k_{g\perp})^2}
+\frac{1}{N_c}\frac{(k_{\perp}-k_{g\perp})\cdot(k_{\perp}-\xi k_{g\perp})}{(k_{\perp}-k_{g\perp})^2(k_{\perp}-\xi k_{g\perp})^2}\Bigg]\non
&-\int^1_{0} d\xi\frac{1+\xi^2}{(1-\xi)_+}x_pq_f(x_p){\cal F}_{x_g}(k_{\perp})\int d^2k_{g\perp}\left[2C_F\frac{1}{k_{g\perp}^2}-N_c\frac{k_{g\perp}\cdot(k_{g\perp}+\xi k_\perp)}{k_{g\perp}^2(k_{g\perp}+\xi k_\perp)^2}+\frac{1}{N_c}\frac{k_{g\perp}\cdot(k_{g\perp}-(1-\xi) k_\perp)}{k_{g\perp}^2(k_{g\perp}-(1-\xi) k_\perp)^2}\right]\Bigg\},
\label{eq:qq-real+virtual}
\end{align}
where the phase space of the radiated gluon has been integrated out. Here we can take the $s\rightarrow\infty$ limit safely. It is straightforward to evaluate most of the terms in the above expression in Eq.~(\ref{eq:qq-real+virtual}) except for the term which is proportional to $\ln(1-\xi)^2$. Following Ref.~\cite{Chirilli:2012jd}, we can make use of the following identity
\begin{align}
\int d^2k_{g\perp}{\cal F}_{x_g}(k_{g\perp})\frac{(k_{\perp}-k_{g\perp})\cdot(k_{\perp}-\xi k_{g\perp})}{(k_{\perp}-k_{g\perp})^2(k_{\perp}-\xi k_{g\perp})^2}
=-\pi {\cal F}_{x_g}(k_\perp)\ln(1-\xi)^2+\pi \overline{I}^{(1)}_{qq}
\end{align}
where $\overline{I}^{(1)}_{qq}$ is defined as
\begin{align}
\overline{I}^{(1)}_{qq}&=\int\frac{d^2k_{g\perp}}{\pi}\left[{\cal F}_{x_g}(k_{g\perp})\frac{(k_\perp-k_{g\perp})\cdot(k_{\perp}-\xi k_{g\perp})}{(k_\perp-k_{g\perp})^2(k_{\perp}-\xi k_{g\perp})^2}
-{\cal F}_{x_g}(k_{\perp})\left\{\frac{(k_\perp-k_{g\perp})\cdot(\xi k_{\perp}-k_{g\perp})}{(k_\perp-k_{g\perp})^2(\xi k_{\perp}-k_{g\perp})^2}
+\frac{k_{g\perp}\cdot(k_{\perp}-k_{g\perp})}{k_{g\perp}^2(k_{\perp}-k_{g\perp})^2}\right\}\right].
\label{eq:Iqq}
\end{align}
Meanwhile, the virtual contribution also contains a similar term in proportion to $\ln(1-\xi)^2$. Then, by combining the real contributions and the virtual contributions together, one finds 
\begin{align}
\int^1_{\frac{\tau}{z}}d\xi\frac{1+\xi^2}{(1-\xi)_+}\frac{x_p}{\xi}q_f\left(\frac{x_p}{\xi}\right)\ln(1-\xi)^2
-\int^1_{0}d\xi\frac{1+\xi^2}{(1-\xi)_+}x_pq_f(x_p)\ln(1-\xi)^2\non
=\int^1_{\frac{\tau}{z}}d\xi\left(\frac{(1+\xi^2)\ln(1-\xi)^2}{1-\xi}\right)_+\frac{x_p}{\xi}q_f\left(\frac{x_p}{\xi}\right).
\end{align}

Regarding the virtual diagrams, each of the virtual diagrams contains UV divergence. 
However, by adding up all of the virtual diagrams, the UV divergences cancel between the virtual contributions. Indeed, we find
\begin{align}
&\int d^2k_{g\perp}\left[2C_F\frac{1}{k_{g\perp}^2}-N_c\frac{k_{g\perp}\cdot(k_{g\perp}+\xi k_\perp)}{k_{g\perp}^2(k_{g\perp}+\xi k_\perp)^2}+\frac{1}{N_c}\frac{k_{g\perp}\cdot(k_{g\perp}-(1-\xi) k_\perp)}{k_{g\perp}^2(k_{g\perp}-(1-\xi) k_\perp)^2}\right]\non
=&\int d^2k_{g\perp}\left[\frac{N_c}{2}\frac{\xi^2 k_\perp^2}{k_{g\perp}^2(k_{g\perp}+\xi k_\perp)^2}-\frac{1}{2N_c}\frac{(1-\xi)^2 k_\perp^2}{k_{g\perp}^2(k_{g\perp}-(1-\xi) k_\perp)^2}\right].
\end{align}

To evaluate the singular part, we need to consider the $\xi$-integral together with the so-called kinematical constraint. As discussed in Ref.~\cite{Mueller:2013wwa, Altinoluk:2014eka,Watanabe:2015tja}, the light cone energy conservation provides the kinematical constraint on $\xi$ as follows
\begin{align}
\xi\le 1-\frac{l_{\perp}^2}{x_ps}.
\label{eq:xi-constraint}
\end{align}
Under this kinematical constraint, the $\xi$-integral gives
\begin{align}
\int_0^{1-\frac{l_\perp^2}{x_ps}} \frac{d\xi}{1-\xi}=\ln\frac{1}{x_g}+\ln\frac{k_{\perp}^2}{l_{\perp}^2}
\label{eq:singularity-terms}
\end{align}
where the first term can be identified as the small-$x$ logarithm, while the second term provides us with an additional power correction. Let us define the rapidity gap between the projectile quark and the target as $Y_g=\ln(1/x_g)$. $Y_g$ goes to infinity when $\sqrt s\rightarrow\infty$, which is known as the rapidity divergence. It can be renormalized into the forward scattering amplitude by employing the BFKL equation at leading-logarithmic accuracy in $\alpha_sY_g$ as follows
\begin{align}
T_{x_g}(x_\perp,y_\perp)=&\;T^{(0)}_{x_g}(x_\perp,y_\perp)\non
&+\frac{\alpha_sN_c}{2\pi^2}Y_g\int d^2b_\perp\frac{(x_\perp-y_\perp)^2}{(x_\perp-b_\perp)^2(b_\perp-y_\perp)^2}
\left[T_{x_g}(x_\perp,b_\perp)+T_{x_g}(b_\perp,y_\perp)-T_{x_g}(x_\perp,y_\perp)\right],
\end{align}
where $T^{(0)}_{x_g}$ can be viewed as the forward scattering amplitude at LO. Next, let us consider the second logarithmic correction in Eq.~(\ref{eq:singularity-terms}). This term does not lead to large contribution when the produced quark and the produced gluon have the same order of the transverse momentum as $|k_\perp|\sim |l_\perp|$. However, we know that this logarithmic correction can be important at high $p_{h\perp}\gtrsim Q_s$~\cite{Watanabe:2015tja}. The corresponding contribution from the real diagrams can be written as 
\begin{align}
\frac{\alpha_sN_c}{2\pi^2}\int_\tau^1\frac{dz}{z^2}D_{h/q}x_pq_f(x_p)\Bigg[&2\int\frac{d^2x_\perp d^2x_\perp^\prime}{(2\pi)^2}e^{-ik_\perp\cdot(x_\perp-x_\perp^\prime)}(-T_{x_g}(x_\perp,x_\perp^\prime))\int d^2l_\perp\frac{1}{l_\perp^2}\ln\frac{k_\perp^2}{l_\perp^2}e^{il_\perp\cdot(x_\perp-x_\perp^\prime)}\non
&-2\int d^2l_\perp\frac{k_\perp\cdot l_\perp}{k_\perp^2l_\perp^2}\ln\frac{k_\perp^2}{l_\perp^2}{\cal F}_{x_g}(k_\perp-l_\perp)\Bigg].
\end{align}
Using the dimensional regularization in the $\overline{\rm MS}$ scheme, the first term in the square bracket yields~\cite{Mueller:2013wwa}
\begin{align}
\int\frac{d^2l_\perp}{l_\perp^2}\ln\frac{k_\perp^2}{l_\perp^2}e^{-il_\perp\cdot r_\perp}=\pi\Bigg[\frac{1}{\epsilon^2}-\frac{1}{\epsilon}\ln\frac{k_\perp^2}{\mu^2}+\frac{1}{2}\left(\ln\frac{k_\perp^2}{\mu^2}\right)^2-\frac{1}{2}\left(\ln\frac{k_\perp^2r_\perp^2}{c_0^2}\right)^2-\frac{\pi^2}{12}\Bigg]
\end{align}
where $c_0=2e^{-\gamma_E}$ with $\gamma_E$ being the Euler constant. In fact, the double pole and single pole vanish by adding another contribution associated with the virtual correction together. One can cast the virtual diagrams with the logarithmic term into 
\begin{align}
-\frac{\alpha_sN_c}{2\pi^2}\int^1_\tau\frac{dz}{z^2}D_{h/q}(z)x_pq_f(x_p){\cal F}_{x_g}(k_\perp)\Bigg[2\int \frac{d^2l_\perp}{l_\perp^2}\ln\frac{k_\perp^2}{l_\perp^2}-2\int d^2l_\perp\frac{l_\perp\cdot(l_\perp+k_\perp)}{l_\perp^2(l_\perp+k_\perp)^2}\ln\frac{k_\perp^2}{l_\perp^2}\Bigg].
\label{eq:power-correction-virtual}
\end{align}

As shown in the Appendix~\ref{AppendixA}, we can cast the above divergent integral into the following form and find
\begin{eqnarray}
&& \int \frac{d^2l_\perp}{(2\pi)^2} \left[\frac{1}{l_\perp^2}-\frac{1}{(l_\perp +k_\perp)^2} +\frac{k_\perp^2}{l_\perp^2 (l_\perp +k_\perp)^2} \right] \ln\frac{k_\perp^2}{l_\perp^2} \notag \\
&=& \frac{1}{2\pi}\left[\frac{1}{\epsilon^2}-\frac{1}{\epsilon}\ln\frac{k_\perp^2}{\mu^2}+\frac{1}{2}\left(\ln\frac{k_\perp^2}{\mu^2}\right)^2-\frac{\pi^2}{12}\right] 
= 2 \int \frac{d^2l_\perp}{(2\pi)^2} \frac{k_\perp^2}{l_\perp^2 (l_\perp +k_\perp)^2} \ln\frac{k_\perp^2}{l_\perp^2}. \label{qqvirtual}
\end{eqnarray} 
After adding the real correction and virtual correction together, we obtain
\begin{align}
\frac{\alpha_s}{2\pi}\int^1_\tau \frac{dz}{z^2}D_{h/q}(z)x_pq_f(x_p)\Bigg[-N_c\int\frac{d^2x_\perp d^2y_\perp}{(2\pi)^2}(-T_{x_g}(x_\perp,y_\perp))e^{-ik_\perp\cdot(x_\perp-y_\perp)}\left(\ln\frac{k_\perp^2(x_\perp-y_\perp)^2}{c_0^2}\right)^2
\non
-\frac{2N_c}{\pi}\int d^2l_\perp \frac{k_\perp\cdot l_\perp}{k_\perp^2l_\perp^2}\ln\frac{k_\perp^2}{l_\perp^2}{\cal F}_{x_g}(k_\perp-l_\perp)\Bigg].
\label{eq:doublelog-correction}
\end{align}
To reach Eq.~(\ref{eq:doublelog-correction}), we have used the same technique as calculations for deriving Sudakov factor~\cite{Mueller:2013wwa}. However, the inclusive hadron production has only one kinematical hard scale, therefore it does not lead to Sudakov factors~\cite{Watanabe:2015tja}.

Finally, let us deal with collinear singularities in Eq.~(\ref{eq:qq-real+virtual}). There are two kinds of collinear singularities: one type of singularity corresponds to the gluon radiation from the incoming quark in the initial state depicted in FIG.~\ref{fig:NLO-qq} $(a)$ and the second one is associated with the final state gluon radiation shown in FIG.~\ref{fig:NLO-qq} $(b)$. In order to extract these collinear singularities from the real diagrams, we use the following identities
\begin{align}
\int d^2k_{g\perp}\frac{1}{(k_\perp-k_{g\perp})^2}{\cal F}_{x_g}(k_{g\perp})
=\;&\pi\int\frac{d^2x_\perp d^2y_\perp}{(2\pi)^2}e^{-ik_\perp\cdot(x_\perp-y_\perp)}(-T_{x_g}(x_\perp,y_\perp))\left(-\frac{1}{\hat\epsilon}+\ln\frac{c_0^2}{\mu^2(x_\perp-y_\perp)^2}\right)
\label{eq:real-collinear-identity1}\\
\int d^2k_{g\perp}\frac{1}{(k_\perp-\xi k_{g\perp})^2}{\cal F}_{x_g}(k_{g\perp})
=\;&\frac{\pi}{\xi^2}\int\frac{d^2x_\perp d^2y_\perp}{(2\pi)^2}e^{-i\frac{k_\perp}{\xi}\cdot(x_\perp-y_\perp)}(-T_{x_g}(x_\perp,y_\perp))\left(-\frac{1}{\hat\epsilon}+\ln\frac{c_0^2}{\mu^2(x_\perp-y_\perp)^2}\right),
\label{eq:real-collinear-identity2}
\end{align}
where the dimensional regularization in the $\overline{\rm MS}$ scheme has been used by setting $1/\hat\epsilon=(4\pi e^{-\gamma_E})^\epsilon/\epsilon$. For the virtual contributions, there are also useful identities 
\begin{align}
\int d^2k_{g\perp}\frac{(\xi k_\perp)^2}{k_{g\perp}^2(k_{g\perp}+\xi k_\perp)^2}&=2\pi\left[-\frac{1}{\hat\epsilon}+\ln\frac{\xi^2k_\perp^2}{\mu^2}\right]
\label{eq:virtual-collinear-identity1}\\
\int d^2k_{g\perp}\frac{((1-\xi) k_\perp)^2}{k_{g\perp}^2(k_{g\perp}-(1-\xi) k_\perp)^2}&=2\pi\left[-\frac{1}{\hat\epsilon}+\ln\frac{(1-\xi)^2k_\perp^2}{\mu^2}\right].
\label{eq:virtual-collinear-identity2}
\end{align}
By adding the LO result, the real contributions and the virtual contributions together, we can absorb the collinear singularity associated with the initial state radiation into the definition of the quark PDF as follows
\begin{align}
q_f(x_p,\mu)=q^{(0)}_f(x_p)-\frac{1}{\hat{\epsilon}}\frac{\alpha_s(\mu)}{2\pi}\int_{\frac{\tau}{z}}^1\frac{d\xi}{\xi} C_F{\cal P}_{qq}(\xi)q_f\left(\frac{x_p}{\xi}\right),
\end{align}
which is exactly the DGLAP evolution equation for the quark PDF. The collinear singularity associated with the final state radiation can be renormalized into the quark FF accordingly
\begin{align}
D_{h/q}(z,\mu)=D_{h/q}^{(0)}(z)-\frac{1}{\hat{\epsilon}}\frac{\alpha_s(\mu)}{2\pi}\int_{z}^1\frac{d\xi}{\xi} C_F{\cal P}_{qq}(\xi)D_{h/q}\left(\frac{z}{\xi}\right),
\end{align}
with ${\cal P}_{qq}(\xi)=\frac{1+\xi^2}{(1-\xi)_+}+\frac{3}{2}\delta(1-\xi)$.

At the end of the day, all the rest of the contributions are finite.  The $q\rightarrow q$ channel contribution of the differential cross section can be written as
\begin{align}
\frac{d\sigma^{{\rm p}+{\rm p}\rightarrow h/q+X}_{(qq)}}{d^2p_{h\perp}dy}=\sum_f\frac{\alpha_s}{2\pi}\int^1_\tau \frac{dz}{z^2}D_{h/q}(z)\int^1_{x_p}\frac{dx}{x}\xi xq_f(x)\int\frac{d^2 x_\perp d^2 y_\perp}{(2\pi)^2}(-T_{x_g}(x_\perp, y_\perp))\left[{\cal H}_{qq}^{(1)}+\int\frac{d^2b_\perp}{(2\pi)^2}{\cal H}_{qq}^{(2)}\right],
\label{eq:qq-final-coordinate}
\end{align}
where the finite hard scattering parts are given by
\begin{align}
{\cal H}_{qq}^{(1)}
=
&\;C_F{\cal P}_{qq}(\xi)\ln\frac{c_0^2}{\mu^2r_\perp^2}\left(e^{-ik_{\perp}\cdot r_\perp}+\frac{1}{\xi^2}e^{-i\frac{k_\perp}{\xi}\cdot r_\perp}\right)
-3C_F\delta(1-\xi)e^{-ik_\perp\cdot r_\perp}\ln\frac{c_0^2}{k_\perp^2r_\perp}\non
&-N_c\delta(1-\xi)e^{-ik_\perp\cdot r_\perp}\int_0^1d\xi^\prime\frac{1+\xi^{\prime2}}{(1-\xi^\prime)_+}\ln\xi^{\prime2}
-\frac{1}{N_c}e^{-ik_\perp\cdot r_\perp}\left(\frac{(1+\xi^2)\ln(1-\xi)^2}{1-\xi}\right)_+\non
&-N_c\delta(1-\xi)e^{-ik_\perp\cdot r_\perp}\left(\ln\frac{k_\perp^2r_\perp^2}{c_0^2}\right)^2
\label{eq:Hqq1}
\end{align}
and
\begin{align}
{\cal H}_{qq}^{(2)}=4\pi\frac{1}{N_c}e^{-ik_\perp\cdot r_\perp}\frac{1+\xi^2}{(1-\xi)_+}I_{qq}^{(1)}+4\pi N_ce^{-ik_\perp\cdot r_\perp}\frac{1+\xi^2}{(1-\xi)_+}I_{qq}^{(2)}-8N_c\pi\delta(1-\xi)\frac{b_\perp\cdot r_\perp}{b_\perp^2 r_\perp^2}\ln\frac{k_\perp^2 r_\perp^2}{c_0^2}e^{-ik_\perp\cdot(b_\perp+r_\perp)}
\label{eq:Hqq2}
\end{align}
with $r_\perp=x_\perp-y_\perp$ and
\begin{align}
I_{qq}^{(1)}&=e^{-i(1-\xi)k_\perp\cdot b_\perp}\left[\frac{b_\perp\cdot(\xi b_\perp-r_\perp)}{b_\perp^2(\xi b_\perp-r_\perp)^2}-\frac{1}{b_\perp^2}\right]+e^{-ik_\perp\cdot b_\perp}\frac{1}{b_\perp^2},\label{eq:I1qq}\\
I_{qq}^{(2)}&=\frac{b_\perp\cdot r_\perp}{b_\perp^2 r_\perp^2}\left[e^{-ik_\perp\cdot b_\perp}+\frac{1}{\xi}e^{-ik_\perp\cdot b_\perp}e^{-i(\frac{1-\xi}{\xi})k_\perp\cdot r_\perp}\right].
\end{align}
In addition, by using the following identities~\cite{Watanabe:2015tja}
\begin{align}
\int\frac{d^2r_\perp}{(2\pi)^2}\ln\frac{c_0^2}{\mu^2r_\perp^2}e^{-ik_\perp\cdot r_\perp}&=\frac{1}{\pi}\left[\frac{1}{k_\perp^2}-2\pi \delta^{(2)}(k_\perp)\int_0^\infty\frac{dl_\perp}{l_\perp}J_0\left(\frac{c_0}{\mu}l_\perp\right)\right],\\
\left(\ln\frac{k_\perp^2r_\perp^2}{c_0^2}\right)^2&=8\pi\int\frac{d^2l_\perp}{(2\pi)^2l_\perp^2}\ln\frac{k_\perp^2}{l_\perp^2}\left[\theta(k_\perp-l_\perp)-e^{-il_\perp\cdot r_\perp}\right],
\end{align}
which gives
\begin{align}
\int\frac{d^2x_\perp d^2y_\perp}{(2\pi)^2}(-T_{x_g}(x_\perp,y_\perp))\ln\frac{c_0^2}{\mu^2r_\perp^2}e^{-ik_\perp\cdot r_\perp}
&=\frac{1}{\pi}\int\frac{d^2l_\perp}{l_\perp^2}\left[{\cal F}_{x_g}(k_\perp+l_\perp)-J_0\left(\frac{c_0}{\mu}l_\perp\right){\cal F}_{x_g}(k_\perp)\right],\\
\int\frac{d^2x_\perp d^2y_\perp}{(2\pi)^2}(-T_{x_g}(x_\perp,y_\perp))\left(\ln\frac{k_\perp^2r_\perp^2}{c_0^2}\right)^2e^{-ik_\perp\cdot r_\perp}
&=\frac{2}{\pi}\int \frac{d^2l_\perp}{l_\perp^2}\ln\frac{k_\perp^2}{l_\perp^2}\left[\theta(k_\perp-l_\perp){\cal F}_{x_g}(k_\perp)-{\cal F}_{x_g}(k_\perp+l_\perp)\right].
\end{align} 
we can cast the NLO corrections into expressions in the momentum space
\begin{align}
\frac{d\sigma^{{\rm p}+{\rm p}\rightarrow h/q+X}_{(qq)}}{d^2p_{h\perp}dy}=\sum_f\frac{\alpha_s}{2\pi}\int^1_\tau \frac{dz}{z^2}D_{h/q}(z)\int^1_{x_p}\frac{dx}{x} \xi xq_f(x)S_{qq},
\label{eq:qq-final-momentum}
\end{align}
where 
\begin{align}
&S_{qq}=\non
&C_F{\cal P}_{qq}(\xi)\frac{1}{\pi}\int\frac{d^2k_{g\perp}}{k_{g\perp}^2}
\Bigg[ {\cal F}_{x_g}(k_\perp+k_{g\perp})-J_0\left(\frac{c_0}{\mu}k_{g\perp}\right){\cal F}_{x_g}(k_\perp)
+\frac{1}{\xi^2}{\cal F}_{x_g}\left(\frac{k_\perp}{\xi}+k_{g\perp}\right)-\frac{1}{\xi^2}J_0\left(\frac{c_0}{\mu}k_{g\perp}\right){\cal F}_{x_g}\left(\frac{k_\perp}{\xi}\right)\Bigg]\non
&-3C_F\delta(1-\xi)\frac{1}{\pi}\int\frac{d^2k_{g\perp}}{k_{g\perp}^2}
\left[{\cal F}_{x_g}(k_\perp+k_{g\perp})-J_0\left(\frac{c_0}{\mu}k_{g\perp}\right){\cal F}_{x_g}(k_\perp)\right]
+N_c\frac{1+\xi^2}{(1-\xi)_+}\overline{I}^{(2)}_{qq}
\non
&
-(2C_F-N_c)\left[\frac{1+\xi^2}{(1-\xi)_+}\overline{I}^{(1)}_{qq}-\left(\frac{(1+\xi^2)\ln(1-\xi)^2}{1-\xi}\right)_+{\cal F}_{x_g}(k_\perp)\right]
-N_c\delta(1-\xi){\cal F}_{x_g}(k_\perp)\int_0^1d\xi^\prime\frac{1+\xi^{\prime2}}{(1-\xi^\prime)_+}\ln \xi^{\prime2}\non
&-\frac{2N_c}{\pi}\delta(1-\xi)\int\frac{d^2k_{g\perp}}{k_{g\perp}^2}\ln\frac{k_\perp^2}{k_{g\perp}^2}\left\{\theta(k_\perp-k_{g\perp}){\cal F}_{x_g}(k_\perp)-{\cal F}_{x_g}(k_\perp-k_{g\perp})\right\}\non
&-\frac{2N_c}{\pi}\delta(1-\xi)\int d^2k_{g\perp} \frac{k_\perp\cdot k_{g\perp}}{k_\perp^2k_{g\perp}^2}\ln\frac{k_\perp^2}{k_{g\perp}^2}{\cal F}_{x_g}(k_\perp-k_{g\perp}),
\end{align}
with
\begin{align}
\overline{I}^{(2)}_{qq}&=\int\frac{d^2k_{g\perp}}{\pi}{\cal F}_{x_g}(k_{g\perp})\left[\frac{k_{\perp}\cdot(k_{g\perp}-k_\perp)}{k_{\perp}^2(k_{g\perp}-k_\perp)^2}+\frac{k_{\perp}\cdot(\xi k_{g\perp}-k_\perp)}{k_{\perp}^2(\xi k_{g\perp}-k_\perp)^2}\right].
\end{align}


\subsection{The $g\rightarrow g$ channel}

\begin{figure}
\centering
\includegraphics[width=16cm,angle=0]{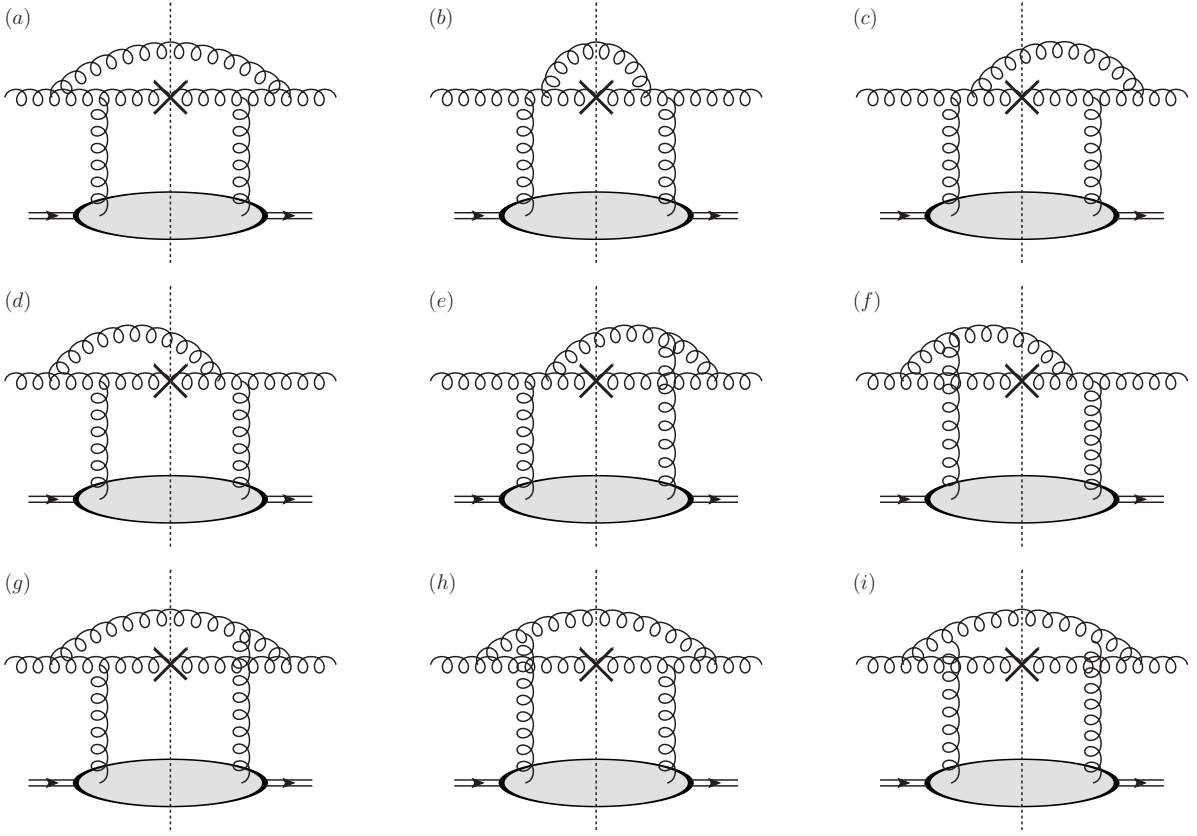}
\caption{Real diagrams at NLO for the $g\rightarrow g$ channel.
 }
\label{fig:NLO-gg-real}
\end{figure}

Now let us consider the $g\rightarrow g$ channel. In the dense regime, the calculations can be more complicated, since the produced two gluons probe higher multi-point Wilson line correlators such as sextupole and quadrupole~\cite{Chirilli:2012jd}, which is suppressed by factors of $N_c$. However, in the dilute regime, we do not have to deal with such problem. For $g+{\rm p}\rightarrow g+g+X$ process shown in FIG.~\ref{fig:NLO-gg-real}, the differential cross section for producing one gluon with $k$ and another gluon $l$ is given by
\begin{align}
&\frac{d\sigma^{g+{\rm p}\rightarrow g+g+X}}{d^3kd^3l}=\alpha_s\delta(p^+-k^+-l^+)\int\frac{d^2x_\perp d^2x_\perp^\prime d^2b_\perp d^2 b_\perp^\prime}{(2\pi)^8}e^{-ik_{\perp}\cdot(x_\perp-x_\perp^\prime)}e^{-il_{\perp}\cdot(b_\perp-b_\perp^\prime)}\sum_{\lambda_1,\lambda_2,\lambda}\psi_{gg\lambda_1\lambda_2}^{\lambda\ast}(u_\perp^\prime)\psi_{gg\lambda_1\lambda_2}^{\lambda}(u_\perp)\non
&\times\Bigg[N_c\widetilde{S}_{x_g}(x_\perp,x_\perp^\prime)+N_c\widetilde{S}_{x_g}(v_\perp,v_\perp^\prime)-\frac{N_c}{2}\widetilde{S}_{x_g}(v_\perp,x_\perp^\prime)-\frac{N_c}{2}\widetilde{S}_{x_g}(v_\perp,b_\perp^\prime)-\frac{N_c}{2}\widetilde{S}_{x_g}(x_\perp,v_\perp^\prime)-\frac{N_c}{2}\widetilde{S}_{x_g}(b_\perp,v_\perp^\prime)\non
&~~~~-\frac{N_c}{2}\widetilde{S}_{x_g}(x_\perp,b_\perp^\prime)-\frac{N_c}{2}\widetilde{S}_{x_g}(b_\perp,x_\perp^\prime)+N_c\widetilde{S}_{x_g}(b_\perp,b_\perp^\prime)\Bigg].
\label{eq:gg-real}
\end{align}
where the color factors are computed by using FeynCalc package~\cite{Shtabovenko:2016sxi}.
The light cone wave function for $g\rightarrow gg$ splitting is defined as~\cite{Dominguez:2011wm}
\begin{align}
\psi^\lambda_{gg\lambda_1\lambda_2}(u_\perp)=\sqrt{\frac{2\xi(1-\xi)}{p^+}}\frac{2\pi i}{u_\perp^2}\Bigg[\frac{1}{\xi}u_\perp\cdot\varepsilon^{(\lambda_1)}_{\perp}\varepsilon^{(\lambda)\ast}_{\perp}\cdot\varepsilon^{(\lambda_2)}_{\perp}
&+\frac{1}{1-\xi}u_\perp\cdot\varepsilon^{(\lambda_2)}_{\perp}\varepsilon^{(\lambda)\ast}_{\perp}\cdot\varepsilon^{(\lambda_1)}_{\perp}
-u_\perp\cdot\varepsilon^{(\lambda)\ast}_{\perp}\varepsilon^{(\lambda_1)}_{\perp}\cdot\varepsilon^{(\lambda_2)}_{\perp}\Bigg]
\end{align}
and, by summing over all of the gluon polarizations, the splitting kernel can be written as
\begin{align}
\sum_{\lambda_1,\lambda_2,\lambda}\psi_{gg\lambda_1\lambda_2}^{\lambda\ast}(u_\perp^\prime)\psi_{gg\lambda_1\lambda_2}^{\lambda}(u_\perp)=\frac{4(2\pi)^2}{p^+}\left[\frac{\xi}{1-\xi}+\frac{1-\xi}{\xi}+\xi(1-\xi)\right]\frac{u_\perp^\prime\cdot u_\perp}{u_\perp^{\prime2}u_\perp^2}.
\end{align}
For inclusive hadron production with the transverse momentum $p_{h\perp}$ at the rapidity $y$, Eq.~(\ref{eq:gg-real}) becomes
\begin{align}
&\frac{\alpha_s}{\pi^2}\int_\tau^1\frac{dz}{z^2}D_{h/g}(z)\int d^2l_\perp \int_{\frac{\tau}{z}}^1d\xi \left[\frac{\xi}{1-\xi}+\frac{1-\xi}{\xi}+\xi(1-\xi)\right]\frac{x_p}{\xi}G\left(\frac{x_p}{\xi}\right)\int\frac{d^2x_\perp d^2x_\perp^\prime d^2b_\perp d^2 b_\perp^\prime}{(2\pi)^4}\non
&\times e^{-ik_{\perp}\cdot(x_\perp-x_\perp^\prime)}e^{-il_{\perp}\cdot(b_\perp-b_\perp^\prime)}\frac{u_\perp^\prime\cdot u_\perp}{u_\perp^{\prime2}u_\perp^2}\Bigg[-N_c\widetilde{T}_{x_g}(x_\perp,x_\perp^\prime)-N_c\widetilde{T}_{x_g}(v_\perp,v_\perp^\prime)+\frac{N_c}{2}\widetilde{T}_{x_g}(v_\perp,x_\perp^\prime)+\frac{N_c}{2}\widetilde{T}_{x_g}(v_\perp,b_\perp^\prime)\non
&~~~~+\frac{N_c}{2}\widetilde{T}_{x_g}(x_\perp,v_\perp^\prime)+\frac{N_c}{2}\widetilde{T}_{x_g}(b_\perp,v_\perp^\prime)+\frac{N_c}{2}\widetilde{T}_{x_g}(x_\perp,b_\perp^\prime)+\frac{N_c}{2}\widetilde{T}_{x_g}(b_\perp,x_\perp^\prime)-N_c\widetilde{T}_{x_g}(b_\perp,b_\perp^\prime)\Bigg].
\label{eq:hadron-gg-real}
\end{align}

\begin{figure}
\centering
\includegraphics[width=18cm,angle=0]{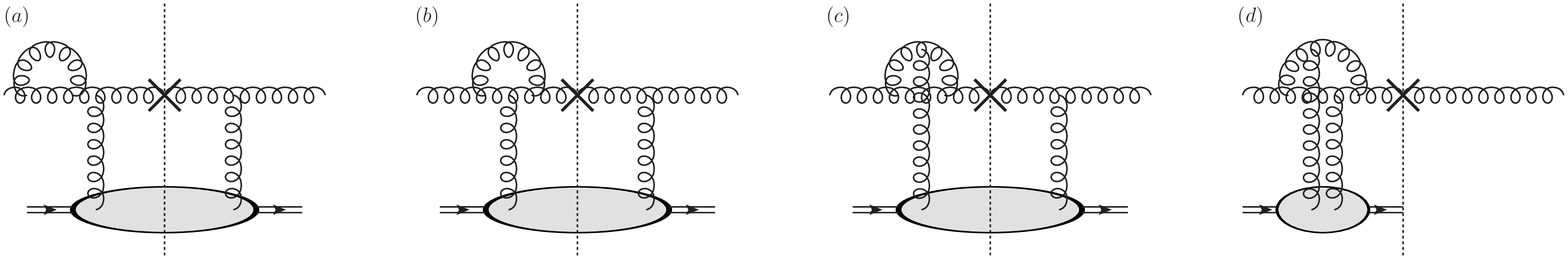}
\caption{Virtual gluon loop corrections at NLO for the $g\rightarrow g$ channel. Mirror contributions are omitted.
 }
\label{fig:NLO-gg-virtual-g}
\end{figure}
\begin{figure}
\centering
\includegraphics[width=18cm,angle=0]{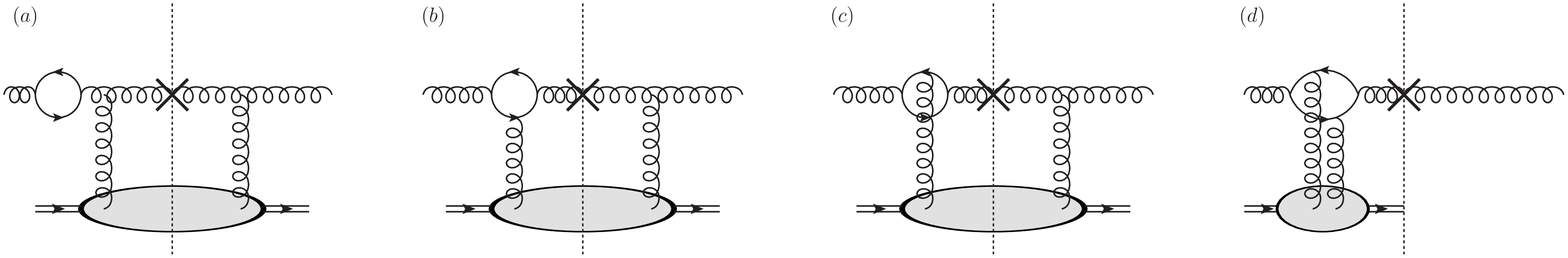}
\caption{Virtual quark loop corrections at NLO for the $g\rightarrow g$ channel. Mirror contributions are omitted.
 }
\label{fig:NLO-gg-virtual-q}
\end{figure}

Furthermore, the gluon virtual contributions (FIG.~\ref{fig:NLO-gg-virtual-g}) yield
\begin{align}
&-\frac{2}{2!}\frac{\alpha_s}{\pi^2}\int^1_\tau\frac{dz}{z^2}D_{h/g}(z)x_pG(x_p)\int^1_0 d\xi \left[\frac{\xi}{1-\xi}+\frac{1-\xi}{\xi}+\xi(1-\xi)\right]\int\frac{d^2v_\perp d^2 v_\perp^\prime d^2u_\perp}{(2\pi)^2}e^{-ik_{\perp}\cdot(v_\perp-v_\perp^\prime)}\frac{1}{u_\perp^2}\non
&\times\Bigg[-N_c\widetilde{T}_{x_g}(v_\perp,v_\perp^\prime)+\frac{N_c}{2}\widetilde{T}_{x_g}(x_\perp,v_\perp^\prime)+\frac{N_c}{2}\widetilde{T}_{x_g}(b_\perp,v_\perp^\prime)+\frac{N_c}{2}\widetilde{T}_{x_g}(x_\perp,b_\perp)\Bigg],
\label{eq:gg-virtual-g}
\end{align}
where the symmetry factor $1/2!$ and a factor of 2 from the mirror diagrams have been taken into consideration. The $\xi$ dependence of Eq.~(\ref{eq:gg-virtual-g}) is symmetry under the interchange $\xi\leftrightarrow 1-\xi$ and the $\xi$ dependent part of the splitting kernel can be rewritten by $2\left[\frac{\xi}{1-\xi}+\frac{1}{2}\xi(1-\xi)\right]$. It is obvious that this splitting function now only contains the rapidity divergence at $\xi=1$. The last diagram in FIG.~\ref{fig:NLO-gg-virtual-g} does not contribute to the inelastic hadron production as is the case for the $q\rightarrow q$ channel, however, the BFKL evolution equation involves this diagram as we will show below.

In addition, we should take into account quark loop corrections (FIG.~\ref{fig:NLO-gg-virtual-q}) which read
\begin{align}
&-2N_f\frac{\alpha_s}{2\pi^2}\int^1_\tau\frac{dz}{z^2}D_{h/g}(z)x_pG(x_p)\int^1_0 d\xi [(1-\xi)^2+\xi^2]\int\frac{d^2v_\perp d^2 v_\perp^\prime d^2u_\perp}{(2\pi)^2}e^{-ik_{\perp}\cdot(v_\perp-v_\perp^\prime)}\frac{1}{u_\perp^2}\non
&\times\Bigg[-T_R\widetilde{T}_{x_g}(v_\perp,v_\perp^\prime)+\frac{T_R}{2}\widetilde{T}_{x_g}(x_\perp,v_\perp^\prime)+\frac{T_R}{2}\widetilde{T}_{x_g}(b_\perp,v_\perp^\prime)+\frac{T_R}{2N_c^2}\widetilde{T}_{x_g}(x_\perp,b_\perp)\Bigg],
\label{eq:gg-virtual-q}
\end{align}
where $T_R={1}/{2}$ for SU(3) and $N_f$ is the number of active flavors in the quark loop. 
A factor of 2 in the front of Eq.~(\ref{eq:gg-virtual-q}) represents the mirror contributions. 
The light cone wave function for $g\rightarrow q\bar q$ splitting, which is used to obtain Eq.~(\ref{eq:gg-virtual-q}), is given by 
\begin{align}
\psi^{\lambda}_{q\bar{q}\alpha\beta}(u_\perp)
=2\pi i\sqrt{\frac{2}{p^+}}
\left\{
\begin{array}{c}
\frac{u_\perp\cdot\varepsilon_\perp^{(1)}}{u_\perp^2}\left[\xi\delta_{\alpha+}\delta_{\beta-}-(1-\xi)\delta_{\alpha-}\delta_{\beta+}\right]~~~~(\lambda=1)\\
\frac{u_\perp\cdot\varepsilon_\perp^{(2)}}{u_\perp^2}\left[\xi\delta_{\alpha-}\delta_{\beta+}-(1-\xi)\delta_{\alpha+}\delta_{\beta-}\right]~~~~(\lambda=2)\\
\end{array}
\right.
\end{align}
and the sum of splitting kernels is
\begin{align}
\sum_{\alpha,\beta,\lambda}\psi_{q\bar q \alpha\beta}^{\lambda\ast}(u_\perp^\prime)\psi_{q\bar q \alpha\beta}^{\lambda}(u_\perp)=\frac{2(2\pi)^2}{p^+}\left[(1-\xi)^2+\xi^2\right]\frac{u_\perp^\prime\cdot u_\perp}{u_\perp^{\prime2}u_\perp^2}.
\label{eq:splitting-kernel-qqbar}
\end{align}
One should keep in mind that there is no singularity in the quark loop diagrams at $\xi\rightarrow1$.

The rapidity divergence in the $g\rightarrow g$ channel can be dealt in the same fashion as in the $q\rightarrow q$ channel.
It is clear that only the real contributions and the gluon virtual loop corrections have the rapidity singularity.
By introducing the plus function as in the $q\rightarrow q$ channel, we can separate the nonsingular part and the singular part. The nonsingular part is manifest
\begin{align}
\frac{\alpha_sN_c}{\pi^2}&\int^1_\tau\frac{dz}{z^2}D_{h/g}(z)\int_{\frac{\tau}{z}}^1d\xi \frac{x_p}{\xi}G\left(\frac{x_p}{\xi}\right)\frac{[1-\xi(1-\xi)]^2}{\xi(1-\xi)_+}
\int d^2k_{g\perp}\widetilde{\cal F}_{x_g}(k_{g\perp})\non
&\times\left[\frac{1}{(k_\perp-k_{g\perp})^2}+\frac{1}{(k_\perp-\xi k_{g\perp})^2}-\frac{k_\perp\cdot(k_\perp-k_{g\perp})}{k_\perp^2(k_\perp-k_{g\perp})^2}-\frac{k_\perp\cdot(k_\perp-\xi k_{g\perp})}{k_\perp^2(k_\perp-\xi k_{g\perp})^2}-\frac{(k_\perp-k_{g\perp})\cdot(k_\perp-\xi k_{g\perp})}{(k_\perp-k_{g\perp})^2(k_\perp-\xi k_{g\perp})^2}\right]\non
-\frac{\alpha_sN_c}{\pi^2}&\int_{\tau}^1\frac{dz}{z^2}D_{h/g}(z)x_pG(x_p)\int_0^1d\xi \left[\frac{\xi}{(1-\xi)_+}+\frac{\xi(1-\xi)}{2}\right]\widetilde{\cal F}_{x_g}(k_\perp)\non
&\times\int d^2k_{g\perp}\frac{1}{2}\left[\frac{(1-\xi)^2k_\perp^2}{k_{g\perp}^2(k_{g\perp}-(1-\xi)k_\perp)^2}+\frac{\xi^2k_\perp^2}{k_{g\perp}^2(k_{g\perp}+\xi k_\perp)^2}\right].
\end{align}
As we mentioned previously, the singular part is also separated the rapidity divergent part from the logarithmic power correction.
Again, the rapidity divergence can be renormalized into the definition of the wave function of the target proton as follows
\begin{align}
\widetilde{T}_{x_g}(x_\perp,y_\perp)=&\;\widetilde{T}^{(0)}_{x_g}(x_\perp,y_\perp)\non
&+\frac{\alpha_sN_c}{2\pi^2}Y_g\int d^2b_\perp\frac{(x_\perp-y_\perp)^2}{(x_\perp-b_\perp)^2(b_\perp-y_\perp)^2}
\left[\widetilde{T}_{x_g}(x_\perp,b_\perp)+\widetilde{T}_{x_g}(b_\perp,y_\perp)-\widetilde{T}_{x_g}(x_\perp,y_\perp)\right]
\end{align}
which is equivalent to the BFKL evolution equation for the fundamental dipole amplitude, since $\widetilde{T}_{x_g}$ can be simply replaced with $N_cT_{x_g}/C_F$ in the dilute regime of the target proton.

The remaining power correction yields
\begin{align}
\frac{\alpha_sN_c}{2\pi^2}&\int_{\tau}^1\frac{dz}{z^2}D_{h/g}(z)x_pG(x_p)\Bigg[2\int\frac{d^2x_\perp d^2y_\perp}{(2\pi)^2}e^{-ik_\perp\cdot r_\perp}(-\widetilde{T}_{x_g}(r_\perp))\int d^2l_\perp\frac{1}{l_\perp^2}\ln\frac{k_\perp^2}{l_\perp^2}e^{il_\perp\cdot r_\perp}\non
&-2\int\frac{d^2x_\perp d^2y_\perp}{(2\pi)^2}e^{-ik_\perp\cdot r_\perp}(-\widetilde{T}_{x_g}(r_\perp))\int d^2l_\perp \frac{k_\perp\cdot l_\perp}{k_\perp^2 l_\perp^2}e^{il_\perp\cdot r_\perp}
\Bigg]\non
-\frac{\alpha_sN_c}{2\pi^2}&\int_{\tau}^1\frac{dz}{z^2}D_{h/g}(z)x_pG(x_p)\widetilde{\cal F}_{x_g}(k_\perp)\left[2\int d^2l_\perp\frac{1}{l_\perp^2}\ln\frac{k_\perp^2}{l_\perp^2}-2\int d^2l_\perp\frac{l_\perp\cdot(l_\perp+k_\perp)}{l_\perp^2(l_\perp+k_\perp)^2}\ln\frac{k_\perp^2}{l_\perp^2}\right].
\label{eq:gg-power-correction}
\end{align}
Then, one finds immediately Eq.~(\ref{eq:gg-power-correction}) can be put into 
\begin{align}
\frac{\alpha_s}{2\pi}\int^1_\tau \frac{dz}{z^2}D_{h/g}(z)x_pG(x_p)\Bigg[-N_c\int\frac{d^2 x_\perp d^2 y_\perp}{(2\pi)^2}(-\widetilde{T}_{x_g}(x_\perp, y_\perp))\left(\ln\frac{k_\perp^2r_\perp^2}{c_0^2}\right)^2e^{-ik_\perp\cdot r_\perp}\non
-\frac{2N_c}{\pi}\int d^2l_\perp \frac{k_\perp\cdot l_\perp}{k_\perp^2l_\perp^2}\ln\frac{k_\perp^2}{l_\perp^2}\widetilde{\cal F}_{x_g}(k_\perp-l_\perp)\Bigg]
\label{eq:doublelog-correction-gg}
\end{align}
with $r_\perp=x_\perp-y_\perp$. In the meantime, the quark virtual corrections is simply given by 
\begin{align}
-\frac{\alpha_sN_fT_R}{2\pi^2}&\int_{\tau}^1\frac{dz}{z^2}D_{h/g}(z)x_pG(x_p)\int_0^1d\xi \left[(1-\xi)^2+\xi^2\right]\non
&\times
\widetilde{\cal F}_{x_g}(k_\perp)\int d^2k_{g\perp}\frac{1}{2}\left[\frac{(1-\xi)^2k_\perp^2}{k_{g\perp}^2(k_{g\perp}-(1-\xi)k_\perp)^2}+\frac{\xi^2k_\perp^2}{k_{g\perp}^2(k_{g\perp}+\xi k_\perp)^2}\right].
\end{align}

Here, the remaining task is to extract the collinear singularities from the real and the virtual contributions, and absorb the pole singularity associated with the initial state radiation into the gluon PDF and another singularity associated with the final state radiation into the gluon FF, respectively. This can be done by using the identities Eqs.~(\ref{eq:real-collinear-identity1})--(\ref{eq:virtual-collinear-identity2}) as follows
\begin{align}
G(x,\mu)&=G^{(0)}(x)-\frac{1}{\hat\epsilon}\frac{\alpha_s(\mu)}{2\pi}\int^1_x\frac{d\xi}{\xi}N_c{\cal P}_{gg}(\xi)G\left(\frac{x}{\xi}\right),\\
D_{h/g}(z,\mu)&=D_{h/g}^{(0)}(z)-\frac{1}{\hat\epsilon}\frac{\alpha_s(\mu)}{2\pi}\int^1_z\frac{d\xi}{\xi}N_c{\cal P}_{gg}(\xi)D_{h/g}\left(\frac{z}{\xi}\right)
\end{align}
where the LO splitting function is defined as
\begin{align}
{\cal P}_{gg}(\xi)=2\left[\frac{\xi}{(1-\xi)_+}+\frac{1-\xi}{\xi}+\xi(1-\xi)\right]+\left(\frac{11}{6}-\frac{2N_fT_R}{3N_c}\right)\delta(1-\xi).
\label{eq:splitting-function-gg}
\end{align}

Finally, for the $g\rightarrow g$ channel, the differential cross section in the coordinate space can be written by
\begin{align}
\frac{d\sigma^{{\rm p}+{\rm p}\rightarrow h/g+X}_{(gg)}}{d^2p_{h\perp}dy}=\frac{\alpha_s}{2\pi}\int^1_\tau\frac{dz}{z^2}D_{h/g}(z)\int^1_{x_p}\frac{dx}{x} \xi xG(x)\int\frac{d^2x_\perp d^2y_\perp}{(2\pi)^2}(-\widetilde{T}_{x_g}(x_\perp,y_\perp))\left[{\cal H}_{gg}^{(1)}+\int\frac{d^2b_\perp}{(2\pi)^2}{\cal H}_{gg}^{(2)}\right]
\label{eq:gg-final-coordinate}
\end{align}
where the hard parts are given by
\begin{align}
{\cal H}_{gg}^{(1)}=~&N_c{\cal P}_{gg}(\xi)\left(e^{-ik_{\perp}\cdot r_\perp}+\frac{1}{\xi^2}e^{-i\frac{k_\perp}{\xi}\cdot r_\perp}\right)\ln\frac{c_0^2}{\mu^2r_\perp^2}
-\left(\frac{11}{3}-\frac{4N_fT_R}{3N_c}\right)N_c\delta(1-\xi)e^{-ik_\perp\cdot r_\perp}\ln\frac{c_0^2}{k_\perp^2r_\perp^2}\non
&-N_c\delta(1-\xi)e^{-ik_\perp\cdot r_\perp}\int^1_0d\xi^\prime 2\left[\frac{\xi^\prime}{(1-\xi^\prime)_+}+\frac{1}{2}\xi^\prime(1-\xi^\prime)\right]\left[\ln\xi^{\prime2}+\ln(1-\xi^\prime)^2\right]\non
&-N_fT_R\delta(1-\xi)e^{-ik_\perp\cdot r_\perp}\int^1_0d\xi^\prime\left[(1-\xi^\prime)^2+\xi^{\prime2}\right]\left[\ln\xi^{\prime2}+\ln(1-\xi^\prime)^2\right]
-N_c\delta(1-\xi)\left(\ln\frac{k_\perp^2r_\perp^2}{c_0^2}\right)^2e^{-ik_\perp\cdot r_\perp},
\label{eq:Hgg1}\\
{\cal H}_{gg}^{(2)}=~&8\pi N_c \frac{[1-\xi(1-\xi)]^2}{\xi(1-\xi)_+}e^{-ik_\perp\cdot r_\perp}\Bigg[\frac{b_\perp\cdot r_\perp}{b_\perp^2r_\perp^2}e^{-ik_\perp\cdot b_\perp}\left(1+\frac{1}{\xi}e^{-i\left(\frac{1-\xi}{\xi}\right)k_\perp\cdot r_\perp}\right)\non
&-\frac{1}{\xi}\frac{(x_\perp-b_\perp)\cdot(y_\perp-b_\perp)}{(x_\perp-b_\perp)^2(y_\perp-b_\perp)^2}e^{-i\left(\frac{1-\xi}{\xi}\right)k_\perp\cdot(b_\perp-y_\perp)}\Bigg]
-8N_c\pi \frac{b_\perp\cdot r_\perp}{b_\perp^2 r_\perp^2}\ln\frac{k_\perp^2 r_\perp^2}{c_0^2}e^{-ik_\perp\cdot(b_\perp+r_\perp)}
\label{eq:Hgg2}
\end{align}
with $r_\perp=x_\perp-y_\perp$.
Through Fourier transform, the differential cross section for the $g\rightarrow g$ channel in the momentum space is 
\begin{align}
\frac{d\sigma^{{\rm p}+{\rm p}\rightarrow h/g+X}_{(gg)}}{d^2p_{h\perp}dy}=\frac{\alpha_s}{2\pi}\int^1_\tau\frac{dz}{z^2}D_{h/g}(z)\int^1_{x_p}\frac{dx}{x} \xi xG(x)S_{gg}
\label{eq:gg-final-momentum}
\end{align}
where 
\begin{align}
S_{gg}=
&\;N_c{\cal P}_{gg}(\xi)\Bigg\{\frac{1}{\pi}\int\frac{d^2k_{g\perp}}{k_{g\perp}}\left[\widetilde{\cal F}_{x_g}(k_{\perp}+k_{g\perp})-J_0\left(\frac{c_0}{\mu}k_{g\perp}\right)\widetilde{\cal F}_{x_g}(k_{\perp})\right]\non
&+\frac{1}{\xi^2}\frac{1}{\pi}\int\frac{d^2k_{g\perp}}{k_{g\perp}}\left[\widetilde{\cal F}_{x_g}\left(\frac{k_{\perp}}{\xi}+k_{g\perp}\right)-J_0\left(\frac{c_0}{\mu}k_{g\perp}\right)\widetilde{\cal F}_{x_g}\left(\frac{k_{\perp}}{\xi}\right)\right]
\Bigg\}\non
&-\left(\frac{11}{3}-\frac{4N_fT_R}{3N_c}\right)N_c\delta(1-\xi)\frac{1}{\pi}\int\frac{d^2k_{g\perp}}{k_{g\perp}}\left[\widetilde{\cal F}_{x_g}(k_{\perp}+k_{g\perp})-J_0\left(\frac{c_0}{k_\perp}k_{g\perp}\right)\widetilde{\cal F}_{x_g}(k_{\perp})\right]\non
&-\delta(1-\xi)\int^1_0d\xi^\prime \left\{N_c\left[\frac{2\xi^\prime}{(1-\xi^\prime)_+}+\xi^\prime(1-\xi^\prime)\right]+N_fT_R\left[(1-\xi^\prime)^2+\xi^{\prime2}\right]\right\}\left[\ln\xi^{\prime2}+\ln(1-\xi^\prime)^2\right]\widetilde{\cal F}_{x_g}(k_\perp)\non
&-8\pi N_c\frac{[1-\xi(1-\xi)]^2}{\xi(1-\xi)_+} \int\frac{d^2k_{g\perp}}{(2\pi)^2}\widetilde{\cal F}_{x_g}(k_{g\perp})\left[\frac{(k_\perp-k_{g\perp})\cdot(k_{\perp}-\xi k_{g\perp})}{(k_\perp-k_{g\perp})^2(k_{\perp}-\xi k_{g\perp})^2}
+\frac{k_\perp\cdot(k_{\perp}-k_{g\perp})}{k_\perp^2(k_{\perp}-k_{g\perp})^2}
+\frac{k_{\perp}\cdot(k_{\perp}-\xi k_{g\perp})}{k_{\perp}^2(k_{\perp}-\xi k_{g\perp})^2}\right]\non
&-\frac{2N_c}{\pi}\delta(1-\xi)\int\frac{d^2k_{g\perp}}{k_{g\perp}^2}\ln\frac{k_\perp^2}{k_{g\perp}^2}\left\{\theta(k_\perp-k_{g\perp})\widetilde{\cal F}_{x_g}(k_\perp)-\widetilde{\cal F}_{x_g}(k_\perp-k_{g\perp})\right\}\non
&-\frac{2N_c}{\pi}\delta(1-\xi)\int d^2k_{g\perp} \frac{k_\perp\cdot k_{g\perp}}{k_\perp^2k_{g\perp}^2}\ln\frac{k_\perp^2}{k_{g\perp}^2}\widetilde{\cal F}_{x_g}(k_\perp-k_{g\perp}).
\end{align}


\subsection{The $q\rightarrow g$ channel}

For the $q\rightarrow g$ channel, there is no virtual correction. The relevant diagrams are the same as in FIG.~\ref{fig:NLO-qq} but the radiated gluon is measured in final state. The differential cross section of ${q+{\rm p}\rightarrow g+q+X}$ for producing a gluon with momentum $k$ and a quark with momentum $l$ is given by
\begin{align}
&\frac{d\sigma^{q+{\rm p}\rightarrow g+q+X}}{d^3kd^3l}
=\;\alpha_s\delta(p^+-k^+-l^+)
\int\frac{d^2x_\perp d^2x_\perp^\prime d^2b_\perp d^2b_\perp^\prime}{(2\pi)^4}
e^{-ik_{\perp}\cdot(x_\perp-x_\perp^\prime)}e^{-il\perp\cdot(b_\perp-b_\perp^\prime)}\sum\limits_{\alpha\beta\lambda}\psi_{gq\alpha\beta}^{\lambda\ast}(u_\perp^\prime)\psi_{gq\alpha\beta}^{\lambda}(u_\perp)\non
\times&
\Bigg[C_FS_{x_g}({x_\perp,x_\perp^\prime})+C_FS_{x_g}({v_\perp,v_\perp^\prime})
-\frac{N_c}{2}S_{x_g}({b_\perp,v_\perp^\prime})-\frac{N_c}{2}S_{x_g}({v_\perp,b_\perp^\prime})-\frac{N_c}{2}S_{x_g}({b_\perp,x_\perp^\prime})-\frac{N_c}{2}S_{x_g}({x_\perp,b_\perp^\prime})\non
&+C_F\widetilde{S}_{x_g}({b_\perp,b_\perp^\prime})+\frac{1}{2N_c}S_{x_g}({x_\perp,v_\perp^\prime})+\frac{1}{2N_c}S_{x_g}({v_\perp,x_\perp^\prime})
\Bigg]
\label{eq:qg-real}
\end{align}
where the splitting kernel is
\begin{align}
\sum\limits_{\alpha\beta\lambda}\psi_{gq\alpha\beta}^{\lambda\ast}(u_\perp^\prime)\psi_{gq\alpha\beta}^{\lambda}(u_\perp)
=\frac{2(2\pi)^2}{p^+}\frac{1+(1-\xi)^2}{\xi}\frac{u_\perp^\prime\cdot u_\perp}{u_\perp^{\prime2}u_\perp^2}.
\end{align}
By convoluting with the quark PDF and the gluon FF, Eq.~(\ref{eq:qg-real}) is cast into
\begin{align}
&\sum_f\frac{\alpha_s}{2\pi^2}\int^1_\tau\frac{dz}{z^2}D_{h/g}(z)\int^1_{\frac{\tau}{z}} d\xi{\cal P}_{gq}(\xi)\frac{x_p}{\xi}q_f\left(\frac{x_p}{\xi}\right)\int d^2k_{g\perp}{\cal F}_{x_g}(k_{g\perp})\Bigg[C_F\frac{1}{(k_{\perp}-\xi k_{g\perp})^2}+C_F\frac{N_c}{C_F}\frac{1}{(k_{g\perp}-k_{\perp})^2}\non
&-N_c\frac{(\xi k_{g\perp}-k_{\perp})\cdot(k_{g\perp}-k_{\perp})}{(\xi k_{g\perp}-k_{\perp})^2(k_{g\perp}-k_{\perp})^2}-N_c\frac{k_{\perp}\cdot(k_{\perp}-k_{g\perp})}{k_{\perp}^2(k_{\perp}-k_{g\perp})^2}+\frac{1}{N_c}\frac{k_{\perp}\cdot(k_{\perp}-\xi k_{g\perp})}{k_{\perp}^2(k_{\perp}-\xi k_{g\perp})^2}
\Bigg]
\label{eq:gq-momentum}
\end{align}
where the splitting function at LO is defined as
\begin{align}
{\cal P}_{gq}(\xi)=\frac{1+(1-\xi)^2}{\xi}.
\label{eq:splitting-function-gq}
\end{align}
The lower limit of the $\xi$-integral is constrained by the kinematics and then the $\xi$ never reaches 0. The remaining task in this channel is to extract the collinear divergences from Eq.~(\ref{eq:gq-momentum}). 
As we mentioned in the previous sections, the collinear divergences associated with the initial state and the final state radiations can be renormalized into the definition of the gluon PDF and the quark FF, respectively
\begin{align}
G(x,\mu)&=G^{(0)}(x)-\frac{1}{\hat\epsilon}\frac{\alpha_s(\mu)}{2\pi}\int^1_{x}\frac{d\xi}{\xi} C_F\sum_f{\cal P}_{gq}(\xi)q_f\left(\frac{x}{\xi}\right)\\
D_{h/q}(z,\mu)&=D_{h/q}^{(0)}(z)-\frac{1}{\hat\epsilon}\frac{\alpha_s(\mu)}{2\pi}\int^1_{z}\frac{d\xi}{\xi} C_F{\cal P}_{gq}(\xi)D_{h/g}\left(\frac{z}{\xi}\right).
\end{align}

In the end, it is easy to obtain the differential cross section in the coordinate space as
\begin{align}
\frac{d\sigma^{{\rm p}+{\rm p}\rightarrow h/g+X}_{(gq)}}{d^2p_{h\perp}dy}
=\sum_f\frac{\alpha_s}{2\pi}\int_\tau^1\frac{dz}{z^2}D_{h/g}(z)\int_{x_p}^1\frac{dx}{x} \xi xq_f(x)\int\frac{d^2x_\perp d^2y_{\perp}}{(2\pi)^2}(-T_{x_g}(x_\perp,y_\perp))\left[{\cal H}^{(1)}_{gq}+\int\frac{d^2b_\perp}{(2\pi)^2}{\cal H}^{(2)}_{gq}\right]
\label{eq:gq-final-coordinate}
\end{align}
where 
\begin{align}
{\cal H}^{(1)}_{gq}=&\;C_F\frac{1}{\xi^2}{\cal P}_{gq}(\xi)e^{-i\frac{k_\perp}{\xi}\cdot r_\perp}\ln\frac{c_0^2}{\mu^2 r_\perp^2}+N_c{\cal P}_{gq}(\xi)e^{-ik_\perp \cdot r_\perp}\ln\frac{c_0^2}{\mu^2 r_\perp^2},
\label{eq:Hgq1}\\
{\cal H}^{(2)}_{gq}=&-4\pi N_c\frac{1}{\xi}{\cal P}_{gq}(\xi)\Bigg[e^{-i\frac{k_\perp}{\xi}\cdot(y_\perp-b_\perp)-ik_\perp\cdot r_\perp}\frac{(b_\perp-y_\perp)\cdot r_\perp}{(b_\perp-y_\perp)^2 r_\perp^2}
+e^{-i\frac{k_\perp}{\xi}\cdot(b_\perp-y_\perp)-ik_\perp\cdot(x_\perp-b_\perp)}\frac{(b_\perp-x_\perp)\cdot(b_\perp-y_\perp)}{(b_\perp-x_\perp)^2(b_\perp-y_\perp)^2}\Bigg]
\non
&
+\frac{4\pi}{N_c}\frac{1}{\xi^2}{\cal P}_{gq}(\xi)e^{-i\frac{k_\perp}{\xi}\cdot r_\perp-i\frac{k_\perp}{\xi}\cdot(y_\perp-b_\perp)}\frac{(b_\perp-y_\perp)\cdot r_\perp}{(b_\perp-y_\perp)^2 r_\perp^2}.
\label{eq:Hgq2}
\end{align}
The last term in ${\cal H}^{(2)}_{gq}$ contributes to inclusive hadron production only in pp collisions but it is subleading compared to the other terms when we take the large-$N_c$. In the momentum space, one obtain easily
\begin{align}
\frac{d\sigma^{{\rm p}+{\rm p}\rightarrow h/g+X}_{(gq)}}{d^2p_{h\perp}dy}
=\sum_f\frac{\alpha_s}{2\pi}\int_\tau^1\frac{dz}{z^2}D_{h/g}(z)\int_{x_p}^1\frac{dx}{x} \xi xq_f(x)S_{gq}
\label{eq:gq-final-momentum}
\end{align}
with
\begin{align}
S_{gq}=
&\;\frac{C_F}{\pi}\frac{1}{\xi^2}{\cal P}_{gq}(\xi)\int \frac{d^2k_{g\perp}}{k_{g\perp}^2}\left[{\cal F}_{x_g}\left(\frac{k_\perp}{\xi}+k_{g\perp}\right)-J_0\left(\frac{c_0}{\mu}k_{g\perp}\right){\cal F}_{x_g}\left(\frac{k_\perp}{\xi}\right)\right]
\non
&
+\frac{N_c}{\pi}{\cal P}_{gq}(\xi)\int \frac{d^2k_{g\perp}}{k_{g\perp}^2}\left[{\cal F}_{x_g}\left(k_\perp+k_{g\perp}\right)-J_0\left(\frac{c_0}{\mu}k_{g\perp}\right){\cal F}_{x_g}\left(k_\perp\right)\right]\non
&-\frac{N_c}{\pi}{\cal P}_{gq}(\xi)\int d^2k_{g\perp}{\cal F}_{x_g}(k_{g\perp})\left[\frac{k_\perp\cdot(k_\perp-k_{g\perp})}{k_\perp^2(k_\perp-k_{g\perp})^2}+\frac{(k_\perp-\xi k_{g\perp})\cdot(k_\perp-k_{g\perp})}{(k_\perp-\xi k_{g\perp})^2(k_\perp-k_{g\perp})^2}\right]
\non
&
+\frac{1}{N_c\pi}{\cal P}_{gq}(\xi)\int d^2k_{g\perp}{\cal F}_{x_g}(k_{g\perp})\frac{k_\perp\cdot(k_\perp-\xi k_{g\perp})}{k_\perp^2(k_\perp-\xi k_{g\perp})^2}.
\end{align}


\subsection{The $g\rightarrow q$ channel}

\begin{figure}
\centering
\includegraphics[width=16cm,angle=0]{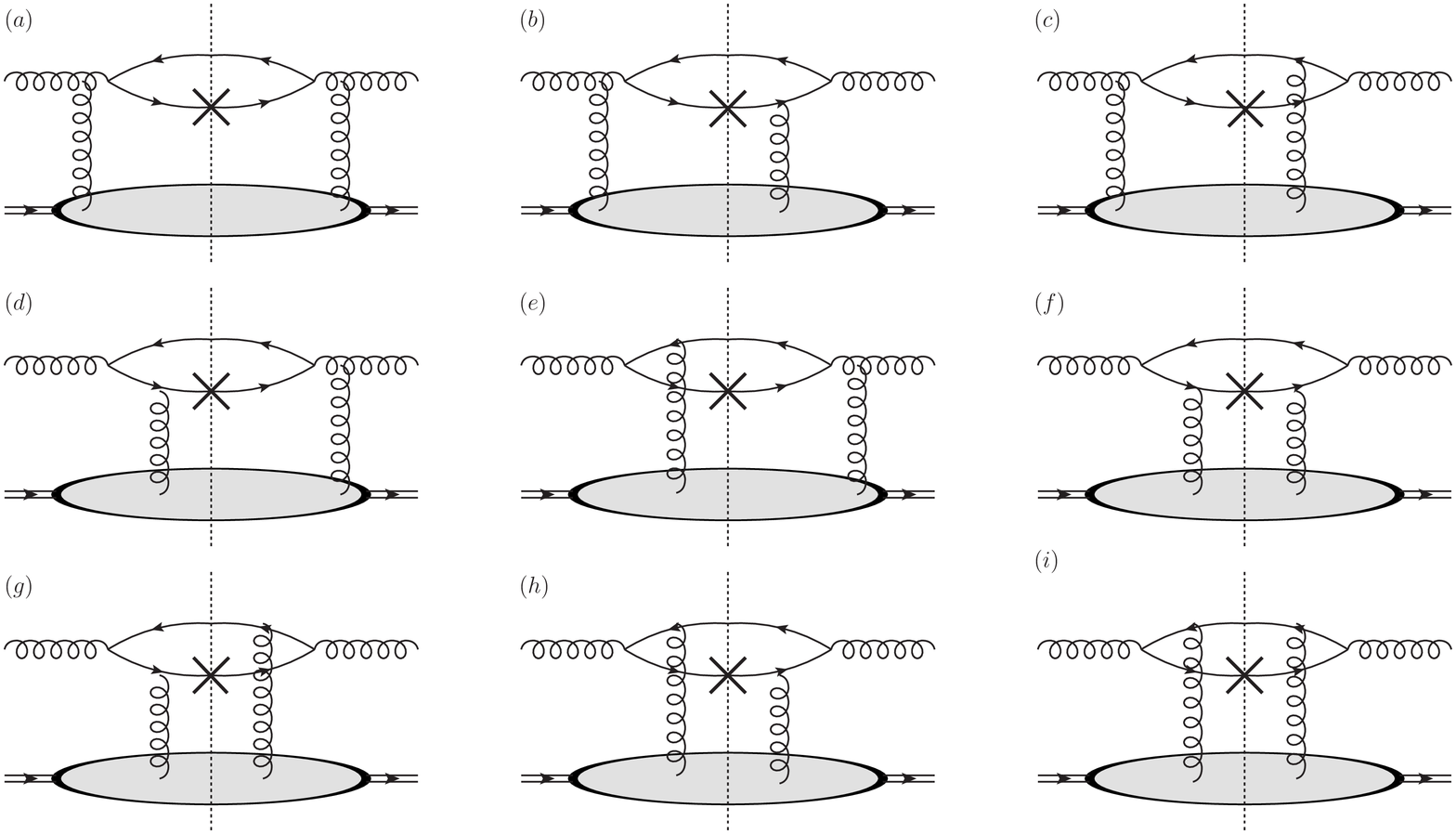}
\caption{Real diagrams at NLO for the $g\rightarrow q$ channel.
 }
\label{fig:NLO-gq-real}
\end{figure}

At last, we consider the $g\rightarrow q$ channel as shown in FIG.~\ref{fig:NLO-gq-real} which has no virtual corrections either. 
The differential cross section of $g+{\rm p}\rightarrow q+\bar q+X$ producing a quark with momentum $k$ and an anti-quark with momentum $l$ is
\begin{align}
\frac{d\sigma^{g+{\rm p}\rightarrow q+\bar q+X}}{d^3kd^3l}=\;&\alpha_s\delta(p^+-k^+-l^+)\int\frac{d^2x_\perp d^2x_\perp^\prime d^2b_\perp d^2b_\perp^\prime}{(2\pi)^8}e^{-ik_{\perp}\cdot(x_\perp-x_\perp^\prime)}e^{-il_{\perp}\cdot(b_\perp-b_\perp^\prime)}\sum_{\alpha,\beta,\lambda}\psi_{q\bar q\alpha\beta}^{\lambda\ast}(u_\perp^\prime)\psi_{q\bar q\alpha\beta}^{\lambda}(u_\perp)\non
&\times\Bigg[T_R\widetilde{S}_{x_g}(v_\perp,v_\perp^\prime)-\frac{T_R}{2}\widetilde{S}_{x_g}(v_\perp,x_\perp^\prime)-\frac{T_R}{2}\widetilde{S}_{x_g}(v_\perp,b_\perp^\prime)-\frac{T_R}{2}\widetilde{S}_{x_g}(x_\perp,v_\perp^\prime)-\frac{T_R}{2}\widetilde{S}_{x_g}(b_\perp,v_\perp^\prime)\non
&+\frac{T_RC_F}{N_c}\widetilde{S}_{x_g}(x_\perp,x_\perp^\prime)+\frac{T_RC_F}{N_c}\widetilde{S}_{x_g}(b_\perp,b_\perp^\prime)+\frac{T_R}{2N_c^2}\widetilde{S}_{x_g}(b_\perp,x_\perp^\prime)+\frac{T_R}{2N_c^2}\widetilde{S}_{x_g}(x_\perp,b_\perp^\prime)\Bigg],
\end{align}
where the splitting kernel is given in Eq.~(\ref{eq:splitting-kernel-qqbar}). By integrating over the phase space of the anti-quark and convoluting with the gluon PDF and the quark FF, one can cast the differential cross section in the momentum space into
\begin{align}
&\frac{\alpha_s}{2\pi^2}T_R\int_\tau^1\frac{dz}{z^2}D_{h/q}(z)\int_{\frac{\tau}{z}}^1d\xi {\cal P}_{qg}(\xi)\frac{x_p}{\xi}G\left(\frac{x_p}{\xi}\right)\int d^2k_{g\perp}{\cal F}_{x_g}(k_{g\perp})\Bigg[\frac{2N_c^2}{d_A}\frac{1}{(k_\perp-\xi k_{g\perp})^2}+\frac{1}{(k_\perp-k_{g\perp})^2}\non
&-\frac{2N_c^2}{d_A}\frac{(k_\perp-k_{g\perp})\cdot(k_\perp-\xi k_{g\perp})}{(k_\perp-k_{g\perp})^2(k_\perp-\xi k_{g\perp})^2}-\frac{2N_c^2}{d_A}\frac{k_\perp\cdot(k_\perp-\xi k_{g\perp})}{k_\perp^2(k_\perp-\xi k_{g\perp})^2}+\frac{2}{d_A}\frac{k_\perp\cdot(k_\perp-k_{g\perp})}{k_\perp^2(k_\perp-k_{g\perp})^2}\Bigg],
\label{eq:qg-momentum}
\end{align}
where $d_A=N_c^2-1$.
The first two terms in the square bracket correspond to the $q\bar q$ splitting after and before the incoming gluon scattering off the target, respectively.
These terms contain the collinear divergences which can be subtracted from Eq.~(\ref{eq:qg-momentum}) and put into the definition of the quark PDF and the gluon FF as follows
\begin{align}
q_f(x,\mu)&=q_f^{(0)}(x)-\frac{1}{\hat\epsilon}\frac{\alpha_s(\mu)}{2\pi}\int_x^1\frac{d\xi}{\xi}T_R{\cal P}_{qg}(\xi)G\left(\frac{x}{\xi}\right),\\
D_{h/g}(z,\mu)&=D_{h/g}^{(0)}(z)-\frac{1}{\hat\epsilon}\frac{\alpha_s(\mu)}{2\pi}\int_z^1\frac{d\xi}{\xi}T_R{\cal P}_{qg}(\xi)D_{h/q}\left(\frac{z}{\xi}\right)
\end{align}
with the LO splitting function
\begin{align}
{\cal P}_{qg}(\xi)=(1-\xi)^2+\xi^2.
\label{eq:splitting-function-qg}
\end{align}
For this channel, caution should be taken when the dimensional regularization is used, since the number of gluon polarization is shifted to $2(1-\epsilon)$. Therefore, the average of gluon polarization yields $\frac{1}{2(1-\epsilon)}$, and thus $-1/\hat\epsilon$ is replaced with $-1/\hat\epsilon-1$ in Eqs.~(\ref{eq:real-collinear-identity1})--(\ref{eq:virtual-collinear-identity2}) eventually. Following the same procedure as shown above, it is straightforward to obtain the differential cross section in the coordinate space
\begin{align}
\frac{d\sigma^{{\rm p}+{\rm p}\rightarrow h/q+X}_{(qg)}}{d^2p_{h\perp}dy}
=&\;\frac{\alpha_s}{2\pi}\int_\tau^1\frac{dz}{z^2}D_{h/q}(z)\int_{x_p}^1\frac{dx}{x} \xi xG\left(x\right)\int\frac{d^2 x_\perp d^2 y_\perp}{(2\pi)^2}(-T_{x_g}(x_\perp,y_\perp))\left[{\cal H}^{(1)}_{qg}+\int\frac{d^2 b_\perp}{(2\pi)^2}{\cal H}^{(2)}_{qg}\right],
\label{eq:qg-final-coordinate}
\end{align}
with the finite hard parts 
\begin{align}
{\cal H}^{(1)}_{qg}=&\;T_R {\cal P}_{qg}(\xi)e^{-ik_{\perp}\cdot r_\perp}\left[\ln\frac{c_0^2}{\mu^2 r_\perp^2}-1\right]
+\frac{1}{\xi^2}\frac{T_R N_c}{C_F}{\cal P}_{qg}(\xi)e^{-i\frac{k_{\perp}}{\xi}\cdot r_\perp}\left[\ln\frac{c_0^2}{\mu^2 r_\perp^2}-1\right],
\label{eq:Hqg1}\\
{\cal H}^{(2)}_{qg}=&-\frac{4\pi}{\xi}\frac{T_RN_c}{C_F}{\cal P}_{qg}(\xi)\Bigg[e^{-i\frac{k_{\perp}}{\xi}\cdot(y_\perp-b_\perp)-ik_{\perp}\cdot(b_\perp-x_\perp)}\frac{(x_\perp-b_\perp)\cdot(y_\perp-b_\perp)}{(x_\perp-b_\perp)^2(y_\perp-b_\perp)^2}+e^{-i\frac{k_{\perp}}{\xi}\cdot r_\perp-ik_{\perp}\cdot(y_\perp-b_\perp)}\frac{(b_\perp-y_\perp)\cdot r_\perp}{(b_\perp-y_\perp)^2 r_\perp^2}\Bigg]\non
&+4\pi \frac{T_R}{N_cC_F}{\cal P}_{qg}(\xi)e^{-ik_{\perp}\cdot r_\perp-ik_{\perp}\cdot(y_\perp-b_\perp)}\frac{(b_\perp-y_\perp)\cdot r_\perp}{(b_\perp-y_\perp)^2 r_\perp^2}.
\label{eq:Hqg2}
\end{align}
The last term in ${\cal H}^{(2)}_{qg}$ is subleading $N_c$ contribution. 
In the momentum space, Eq.~(\ref{eq:qg-final-coordinate}) becomes
\begin{align}
\frac{d\sigma^{{\rm p}+{\rm p}\rightarrow h/q+X}_{(qg)}}{d^2p_{h\perp}dy}
=\frac{\alpha_s}{2\pi}\int_\tau^1\frac{dz}{z^2}D_{h/q}(z)\int_{x_p}^1\frac{dx}{x} \xi xG\left(x\right)S_{qg},
\label{eq:qg-final-momentum}
\end{align}
with  
\begin{align}
S_{qg}&=T_R{\cal P}_{qg}(\xi)\left\{\frac{1}{\pi}\int\frac{d^2k_{g\perp}}{k_{g\perp}^2}\left[{\cal F}_{x_g}(k_{\perp}+k_{g\perp})-J_0\left(\frac{c_0}{\mu}k_{g\perp}\right){\cal F}_{x_g}(k_{\perp})\right]-{\cal F}_{x_g}(k_{\perp})\right\}\non
&+\frac{1}{\xi^2}\frac{T_RN_c}{C_F}{\cal P}_{qg}(\xi)\left\{\frac{1}{\pi}\int\frac{d^2k_{g\perp}}{k_{g\perp}^2}\left[{\cal F}_{x_g}\left(\frac{k_{\perp}}{\xi}+k_{g\perp}\right)-J_0\left(\frac{c_0}{\mu}k_{g\perp}\right){\cal F}_{x_g}\left(\frac{k_{\perp}}{\xi}\right)\right]-{\cal F}_{x_g}\left(\frac{k_{\perp}}{\xi}\right)\right\}\non
&-\frac{T_RN_c}{\pi C_F}{\cal P}_{qg}(\xi)\int d^2k_{g\perp}{\cal F}_{x_g}(k_{g\perp})\left[\frac{(k_{\perp}-k_{g\perp})\cdot(k_{\perp}-\xi k_{g\perp})}{(k_{\perp}-k_{g\perp})^2(k_{\perp}-\xi k_{g\perp})^2}+\frac{k_{\perp}\cdot(k_{\perp}-\xi k_{g\perp})}{k_{\perp}^2(k_{\perp}-\xi k_{g\perp})^2}\right]\non
&+\frac{T_R}{\pi N_cC_F}{\cal P}_{qg}(\xi)\int d^2k_{g\perp}{\cal F}_{x_g}(k_{g\perp})\frac{k_{\perp}\cdot(k_{\perp}-k_{g\perp})}{k_{\perp}^2(k_{\perp}-k_{g\perp})^2}.
\end{align}


\section{Summary}

In this paper, we have calculated the inclusive hadron production at forward rapidity in pp collisions in the dilute regime with finite $N_c$. 
Using the dimensional regularization with the $\overline{\rm MS}$ scheme, we have shown that the collinear divergences and the rapidity divergences can be separated from the hard scattering parts and renormalized into the PDFs, the FFs, or the wave function of the target proton.

Here, let us summarize all of the results by adding up the LO differential cross section and all four channels of the NLO corrections together. In the momentum space, the master formula can be written as
\begin{align}
\frac{d\sigma^{{\rm p}+{\rm p}\rightarrow h+X}}{d^2p_{h\perp}dy}
=\sum_f\int_{\tau}^1\frac{dz}{z^2}\int_{x_p}^1\frac{dx}{x}\xi
\left(xq_f\left(x,\mu\right),xG\left(x,\mu\right)\right)
\begin{pmatrix}
S^{(0)}_{qq}+\frac{\alpha_s}{2\pi}S_{qq} & \frac{\alpha_s}{2\pi}S_{gq}\\
\frac{\alpha_s}{2\pi}S_{qg} & S^{(0)}_{gg}+\frac{\alpha_s}{2\pi}S_{gg}
\end{pmatrix}
\begin{pmatrix}
D_{h/q}(z,\mu)\\
D_{h/g}(z,\mu)
\end{pmatrix}
\label{eq:master-expression}
\end{align}
where $S_{qq}^{(0)}=\delta(1-\xi){\cal F}_{x_g}(k_\perp)$ and $S_{gg}^{(0)}=\delta(1-\xi)\widetilde{\cal F}_{x_g}(k_\perp)$.
As shown in the Appendix~\ref{AppendixB}, basically, the hard coefficients for all four channels derived in this paper are equivalent to those obtained in p$A$ collisions in Ref.~\cite{Chirilli:2012jd} after taking the dilute limit as well as the large $N_c$ limit.  

The PDFs depend on the factorization scale and obey the DGLAP evolution equation
\begin{align}
\begin{pmatrix}
q_f(x,\mu)\\
G(x,\mu)
\end{pmatrix}
=
\begin{pmatrix}
q_f^{(0)}(x)\\
G^{(0)}(x)
\end{pmatrix}
-\frac{1}{\hat\epsilon}\frac{\alpha_s(\mu)}{2\pi}\int_{z}^1\frac{d\xi}{\xi}
\begin{pmatrix}
C_F{\cal P}_{qq}(\xi)&T_R{\cal P}_{qg}(\xi)\\
\sum_fC_F{\cal P}_{gq}(\xi)&N_c{\cal P}_{gg}(\xi)
\end{pmatrix}
\begin{pmatrix}
q_f(x/\xi)\\
G(x/\xi)
\end{pmatrix}
.
\label{eq:DGLAP-PDF}
\end{align}
Similarly, the FFs obey
\begin{align}
\begin{pmatrix}
D_{h/q}(z,\mu)\\
D_{h/g}(z,\mu)
\end{pmatrix}
=
\begin{pmatrix}
D^{(0)}_{h/q}(z)\\
D^{(0)}_{h/g}(z)
\end{pmatrix}
-\frac{1}{\hat\epsilon}\frac{\alpha_s(\mu)}{2\pi}\int_{z}^1\frac{d\xi}{\xi}
\begin{pmatrix}
C_F{\cal P}_{qq}(\xi)&C_F{\cal P}_{gq}(\xi)\\
\sum_fT_R{\cal P}_{qg}(\xi)&N_c{\cal P}_{gg}(\xi)
\end{pmatrix}
\begin{pmatrix}
D_{h/q}(z/\xi)\\
D_{h/g}(z/\xi)
\end{pmatrix}
.
\label{eq:DGLAP-FF}
\end{align}
The rapidity divergences can be dealt with the BFKL equation. In the leading logarithmic approximation, the BFKL equation can be written as 
\begin{align}
\frac{\partial T_{x_g}(r_\perp)}{\partial Y_g}=
\frac{\overline{\alpha}_s}{2\pi}\int d^2b_\perp\frac{r_\perp^2}{(r_\perp-b_\perp)^2b_\perp^2}
\left[T_{x_g}(r_\perp-b_\perp)+T_{x_g}(b_\perp)-T_{x_g}(r_\perp)\right],
\end{align}
with $\overline{\alpha}_s=\alpha_s N_c/\pi$. The factorization scale $\mu$ dependence is largely cancelled between the the hard parts and the PDFs and the FFs. Putting in the one-loop running coupling constant in the hard parts and using the NLO DGLAP equation and the NLO BFKL equation, Eq.~(\ref{eq:master-expression}) provides the complete results at NLO for inclusive hadron production in pp collisions in the dilute regime.

It would be very interesting to evaluate Eq.~(\ref{eq:master-expression}) numerically for pp and p$A$ collisions and compare with the results obtained from the non-linear formulation~\cite{Watanabe:2015tja}. As to the initial condition for the BFKL evolution, we could use either use the dipole-dipole scattering amplitude, or models that we used for the heavy nuclei such as the McLerran-Venugopalan~\cite{McLerran:1993ni} or GBW model~\cite{GolecBiernat:1998js} in the dilute regime.
Nevertheless, once the same initial conditions are set for both the BFKL equation and the BK equation, we anticipate that Eq.~(\ref{eq:master-expression}) can reveal a difference between the small-$x$ formalism in pp and p$A$ collisions for hadron spectra in the low transverse momentum region. This interesting comparison can provides us the precise and important information about the role of the non-linear gluon dynamics in these collisions, and thus help us observe the onset the saturation effects. We leave this issue for future study.

\section*{Acknowledgements}
The authors are grateful to A.~H.~Mueller and F.~Yuan for useful discussions. This work is supported by the NSFC under Grant No.~11575070.

\appendix
\section{Evaluation of Several Integrals in Dimensional Regularization}{\label{AppendixA}}
In this part, we provide some technical details for the evaluation of several divergent integrals in dimensional regularization. The most important step in the evaluation of Eq.~(\ref{qqvirtual}) is to see that it is related to the function $I(a)$ which is defined as 
\begin{equation}
I(a) =\int \frac{d^2l_\perp}{(2\pi)^2} \left[\frac{1}{l_\perp^2}-\frac{1}{(l_\perp +k_\perp)^2} +\frac{k_\perp^2}{l_\perp^2 (l_\perp +k_\perp)^2} \right] \left(\frac{k_\perp^2}{l_\perp^2}\right)^a. \label{gen}
\end{equation}
It is straightforward to find that the integral in Eq.~(\ref{qqvirtual}) is given by $\left. \frac{\partial I(a)}{\partial a}\right|_{a=0}$. Furthermore, it is also important to note that Eq.~(\ref{gen}) contains only IR divergence at $l_\perp =0$, while it does not have UV divergence when $l_\perp \to \infty$ and it is finite when $l_\perp +k_\perp =0$. In addition, having dimensional regularization in mind, we only have to evaluate the last two terms in the square brackets of Eq.~(\ref{gen}) since the first term is identically zero in dimensional regularization. Therefore, in order to get $I(a)$, we only need to consider the following integral
\begin{align}
J(a)=\int \frac{d^{2}l_\perp}{(2\pi)^{2}}\frac{1}{(l_\perp+k_{\perp})^2}
\left(\frac{k_{\perp}^2}{l_\perp^2}\right)^a, 
\label{eq;main-integral}
\end{align}
since the last term inside the square brackets of Eq.~(\ref{gen}) can be viewed as $J(a+1)$. Using the following identity, 
\begin{align}	
\frac{1}{(q_\perp^{2})^a}=\frac{1}{\Gamma(a)}\int_0^\infty dx x^{a-1}e^{-xq_\perp^2},
\end{align}
and adopting the dimensional regularization in the modified minimal subtraction ($\overline{\textrm{MS}}$) scheme by shifting the dimension of $l_\perp$ integration ($2\rightarrow 2-2\epsilon$) , we find that Eq.~(\ref{eq;main-integral}) can be cast into
\begin{align}
\left(\frac{\mu^{2}e^{\gamma_E}}{4\pi}\right)^{\epsilon}\int \frac{d^{2-2\epsilon}l_\perp}{(2\pi)^{2-2\epsilon}}k_{\perp}^{2a}
\frac{1}{\Gamma(a)}
\int^\infty_0dxdy e^{-(x+y)\left(l_\perp+\frac{x}{x+y}k_{\perp}\right)^2-\frac{xy}{x+y}k_{\perp}^2},
\label{eq;integral2}
\end{align}
with $\gamma_E$ the Euler constant. After changing variable $l_\perp^\prime=l_\perp+\frac{x}{x+y}k_{\perp}$ and $t=y/(x+y)$, it is straightforward to find
\begin{equation}
J(a) \overset{\overline{\text{MS}}}{=\joinrel=}
\frac{1}{4\pi}\left(\frac{e^{\gamma_E}\mu^2}{k_{\perp}^2}\right)^\epsilon
\frac{\Gamma(\epsilon+a)}{\Gamma(a)}
\frac{\Gamma(-\epsilon)\Gamma(-\epsilon-a+1)}{\Gamma(-2\epsilon-a+1)}.
\label{eq;final-result}
\end{equation}
Using Eq.~(\ref{eq;final-result}) and the trick of differentiation mentioned above, we can obtain 
\begin{equation}
\int \frac{d^2l_\perp}{(2\pi)^2} \frac{k_\perp^2}{l_\perp^2 (l_\perp +k_\perp)^2} \ln\frac{k_\perp^2}{l_\perp^2}=\left. \frac{\partial J(a+1)}{\partial a}\right|_{a=0}=\frac{1}{4\pi}\left(\frac{1}{\epsilon^2}-\frac{1}{\epsilon}\ln\frac{k_{\perp}^2}{\mu^2}-\frac{\pi^2}{12}+\frac{1}{2}\ln^2\frac{k_{\perp}^2}{\mu^2}\right).
\end{equation}
In addition, in dimensional regularization, we can also find 
\begin{equation}
I(a) = J(a+1) -J(a)=\frac{1}{2\pi}\left(\frac{e^{\gamma_E}\mu^2}{k_{\perp}^2}\right)^\epsilon
\frac{\Gamma(\epsilon+a+1)}{\Gamma(a+1)}
\frac{\Gamma(1 -\epsilon)\Gamma(-\epsilon-a)}{\Gamma(1-2\epsilon-a)}.
\end{equation}
At last, by taking the derivative of $I(a)$ with respect to $a$ at $a=0$, we can obtain 
\begin{align}
\int \frac{d^2l_\perp}{(2\pi)^2} \left[\frac{1}{l_\perp^2}-\frac{1}{(l_\perp +k_\perp)^2} +\frac{k_\perp^2}{l_\perp^2 (l_\perp +k_\perp)^2} \right] \ln \frac{k_\perp^2}{l_\perp^2}
\overset{\overline{\text{MS}}}{=\joinrel=}&~
\frac{1}{2\pi}\left(\frac{1}{\epsilon^2}-\frac{1}{\epsilon}\ln\frac{k_{\perp}^2}{\mu^2}-\frac{\pi^2}{12}+\frac{1}{2}\ln^2\frac{k_{\perp}^2}{\mu^2}\right), 
\label{eq:dim-regularization-log-integral}
\end{align}
which gives the results used in Eq.~(\ref{qqvirtual}). As mentioned in Ref.~\cite{Mueller:2015ael}, there is a very quick way to derive the complete virtual contribution based on the observation that the sum of virtual contributions is free of UV divergence while only the self-energy diagram contains IR divergence. Therefore, we can obtain the full virtual contribution by simply putting a UV cutoff $k_\perp$ on the self-energy contribution as follows
\begin{equation}
2\left. \int\frac{d^2 l_\perp}{(2\pi)^2}\frac{1}{l_\perp^2}\ln\frac{k_\perp^2}{l_\perp^2}\right|_{l_\perp<k_\perp}
=
\frac{1}{2\pi}\left[\frac{1}{\epsilon^2}-\frac{1}{\epsilon}\ln\frac{k_\perp^2}{\mu^2}
+\frac{1}{2}\ln^2\frac{k_\perp^2}{\mu^2}
-\frac{\pi^2}{12}\right] ,\label{vircq}
\end{equation}
which is identical to the results in Eq.~(\ref{eq:dim-regularization-log-integral}).

\section{Comparing our results with the non-linear results in p$A$ collisions}{\label{AppendixB}}

We linearize the results of inclusive hadron production cross section in p$A$ collisions in the small-$x$ saturation formalism derived in Refs.~\cite{Chirilli:2012jd, Watanabe:2015tja}. This procedure can provide a useful cross check of the results obtained in this paper and also reveal a difference between the small-$x$ formalism in pp and p$A$ collisions. 
As shown in Ref.~\cite{Chirilli:2012jd}, the collinear divergences can be renormalized into the PDFs and the FFs and the rapidity divergence is dealt with the BK equation. Therefore, we should look at the hard coefficients only. 

\subsection{The $q\rightarrow q$ channel}

Let us consider the original result as shown in Eq.~(41) in Ref.~\cite{Chirilli:2012jd} 
\begin{align}
\frac{d\sigma_{(qq)}^{{\rm p}+A\rightarrow h/q+X}}{d^2p_{h\perp}dy}
=\sum_f\int_{\tau}^1\frac{dz}{z^2}\int_{x_p}^1\frac{dx}{x}D_{h/q}\xi xq_f(x)\frac{\alpha_s}{2\pi}
\int\frac{d^2x_\perp d^2y_\perp}{(2\pi)^2}\left[S_{x_g}(x_\perp,y_\perp){\cal H}_{2qq}^{(1)}
+\int\frac{d^2b_\perp}{(2\pi)^2}S_{x_g}^{(4)}(x_\perp,b_\perp,y_\perp){\cal H}_{4qq}^{(1)}
\right]
\label{eq:CXY-41}
\end{align}
where the hard coefficients are
\begin{align}
{\cal H}_{2qq}^{(1)}=\;
&C_F{\cal P}_{qq}(\xi)\ln\frac{c_0^2}{\mu^2r_\perp^2}\left(e^{-ik_{\perp}\cdot r_\perp}+\frac{1}{\xi^2}e^{-i\frac{k_\perp}{\xi}\cdot r_\perp}\right)
-3C_F\delta(1-\xi)e^{-ik_\perp\cdot r_\perp}\ln\frac{c_0^2}{k_\perp^2r_\perp}\non
&-(2C_F-N_c)e^{-ik_\perp\cdot r_\perp}\left[\frac{1+\xi^2}{(1-\xi)_+}\tilde{I}_{21}-\left(\frac{(1+\xi^2)\ln(1-\xi)^2}{1-\xi}\right)_+\right],
\\
{\cal H}_{4qq}^{(1)}=&-4\pi N_ce^{-ik_\perp\cdot r_\perp}\Bigg[e^{-i\frac{1-\xi}{\xi}k_\perp\cdot(x_\perp-b_\perp)}\frac{1+\xi^2}{(1-\xi)_+}\frac{1}{\xi}\frac{(x_\perp-b_\perp)\cdot (y_\perp-b_\perp)}{(x_\perp-b_\perp)^2(y_\perp-b_\perp)^2}\non
&-\delta(1-\xi)\int_0^1d\xi^\prime\frac{1+\xi^{\prime2}}{(1-\xi)_+}\Bigg[\frac{e^{-i(1-\xi^\prime)k_\perp\cdot(y_\perp)}}{(b_\perp-y_\perp)^2}
-\delta^{(2)}(b_\perp-y_\perp)\int d^2r_\perp^\prime \frac{e^{ik_\perp\cdot r_\perp^\prime}}{r_\perp^{\prime2}}\Bigg]
\Bigg]
\end{align}
with $\tilde{I}_{21}=\int\frac{d^2b_\perp}{\pi}I^{(1)}_{qq}$ and $I^{(1)}_{qq}$ is given in Eq.~(\ref{eq:I1qq}).
The quadrupole amplitude is defined as
\begin{align}
S_{x_g}^{(4)}(x_\perp,b_\perp,y_\perp)=\frac{1}{N_c^2}\langle {\rm Tr}\left[U(x_\perp)U^\dagger(b_\perp)\right] {\rm Tr}\left[U(b_\perp)U^\dagger(y_\perp)\right]\rangle_{x_g}
\simeq S_{x_g}(x_\perp,b_\perp)S_{x_g}(b_\perp,y_\perp)
\end{align}
where the last line is valid only in the large-$N_c$ limit. 
By expanding the quadrupole amplitude in the $T_{x_g}$, Eq.~(\ref{eq:CXY-41}) can be cast into
\begin{align}
\sum_f\int_{\tau}^1\frac{dz}{z^2}\int_{x_p}^1\frac{dx}{x}D_{h/q}\xi xq_f(x)\frac{\alpha_s}{2\pi}\int\frac{d^2x_\perp d^2y_\perp}{(2\pi)^2}(-T_{x_g}(r_\perp))\left[{\mathbf H}_{2qq}^{(1)}+\int\frac{d^2b_\perp}{(2\pi)^2}{\mathbf H}_{2qq}^{(2)}\right]
\end{align}
where the hard coefficients are 
\begin{align}
{\mathbf H}_{2qq}^{(1)}=&\;
C_F{\cal P}_{qq}(\xi)\ln\frac{c_0^2}{\mu^2r_\perp^2}\left(e^{-ik_{\perp}\cdot r_\perp}+\frac{1}{\xi^2}e^{-i\frac{k_\perp}{\xi}\cdot r_\perp}\right)
-3C_F\delta(1-\xi)e^{-ik_\perp\cdot r_\perp}\ln\frac{c_0^2}{k_\perp^2r_\perp}\non
&-\frac{1}{N_c}e^{-ik_\perp\cdot r_\perp}\left(\frac{(1+\xi^2)\ln(1-\xi)^2}{1-\xi}\right)_+
-N_c\delta(1-\xi)e^{-ik_\perp\cdot r_\perp}\int_0^1d\xi^\prime\frac{1+\xi^{\prime2}}{(1-\xi^\prime)_+}\ln\xi^{\prime2},
\\
{\mathbf H}_{2qq}^{(2)}=&\;
4\pi \frac{1}{N_c}e^{-ik_\perp\cdot r_\perp}\frac{1+\xi^2}{(1-\xi)_+}I^{(1)}_{qq}
+4\pi N_ce^{-ik_\perp\cdot r_\perp}\frac{1+\xi^2}{(1-\xi)_+}I^{(2)}_{qq}.
\end{align}
Then, by replacing the transverse area of target nucleus with that of proton, one finds ${\mathbf H}_{2qq}^{(1)}$ is equivalent to Eq.~(\ref{eq:Hqq1}) except for the double logarithmic term and ${\mathbf H}_{2qq}^{(2)}$ matches Eq.~(\ref{eq:Hqq2}) except for the last term. The remaining power corrections in Eqs.~(\ref{eq:Hqq1})(\ref{eq:Hqq2}) can be derived from $L_q$ term as we will show below.

\subsection{The $g\rightarrow g$ channel}

For the $g\rightarrow g$ channel, one finds the final results in Eq.~(74) in Ref.~\cite{Chirilli:2012jd} 
\begin{align}
\frac{d\sigma_{(gg)}^{{\rm p}+A\rightarrow h/g+X}}{d^2p_{h\perp}dy}
=&\int_{\tau}^1\frac{dz}{z^2}\int_{x_p}^1\frac{dx}{x}D_{h/g}\xi xG(x)\frac{\alpha_s}{2\pi}\int\frac{d^2x_\perp d^2y_\perp}{(2\pi)^2}\Bigg[
S_{x_g}(x_\perp,y_\perp)S_{x_g}(y_\perp,x_\perp){\cal H}_{2gg}^{(1)}\non
&+\int\frac{d^2b_\perp}{(2\pi)^2}\left\{
S_{x_g}(x_\perp,b_\perp)S_{x_g}(b_\perp,y_\perp){\cal H}_{2q\bar{q}}^{(1)}
+
S_{x_g}(x_\perp,b_\perp)S_{x_g}(b_\perp,y_\perp)S_{x_g}(y_\perp,x_\perp){\cal H}_{6gg}^{(1)}
\right\}
\Bigg]
\label{eq:CXY-74}
\end{align}
with
\begin{align}
{\cal H}_{2gg}^{(1)}=&\;N_c{\cal P}_{gg}(\xi)\ln\frac{c_0^2}{r_\perp^2\mu^2}\left(e^{-ik_\perp\cdot r_\perp}+\frac{1}{\xi^2}e^{-i\frac{k_\perp}{\xi}\cdot r_\perp}\right)-\left(\frac{11}{3}-\frac{4N_fT_R}{3N_c}\right)N_c\delta(1-\xi)e^{-ik_\perp\cdot r_\perp}\ln\frac{c_0^2}{r_\perp^2k_\perp^2},\\
{\cal H}_{2q\bar q}^{(1)}=&\;8\pi N_fT_Re^{-ik_\perp\cdot (y_\perp-b_\perp)}\delta(1-\xi)\int_0^1d\xi^\prime\left[\xi^{\prime2}+(1-\xi^\prime)^2\right]\left[\frac{e^{-i\xi^\prime k_\perp\cdot r_\perp}}{r_\perp^2}-\delta^{(2)}(r_\perp)\int d^2r_\perp^\prime \frac{e^{ik_\perp\cdot r_\perp^\prime}}{r_\perp^{\prime2}}\right],\\
{\cal H}_{6gg}^{(1)}=&-16\pi N_c e^{-ik_\perp\cdot r_\perp}\Bigg\{e^{-i\frac{k_\perp}{\xi}\cdot(y_\perp-b_\perp)}\frac{[1-\xi(1-\xi)]^2}{(1-\xi)_+}\frac{1}{\xi^2}\frac{r_\perp\cdot (b_\perp-y_\perp)}{r_\perp^2(b_\perp-y_\perp)^2}\non
&-\delta(1-\xi)\int_0^1d\xi^\prime\left[\frac{\xi^\prime}{(1-\xi^\prime)_+}+\frac{1}{2}\xi^\prime(1-\xi^\prime)\right]\left[\frac{e^{-i\xi^\prime k_\perp\cdot (y_\perp-b_\perp)}}{(b_\perp-y_\perp)^2}-\delta^{(2)}(b_\perp-y_\perp)\int d^2r_\perp^\prime\frac{e^{ik_\perp\cdot r_\perp^{\prime}}}{r_\perp^{\prime2}}\right]
\Bigg\}.
\end{align}
where the large-$N_c$ limit is taken in the derivation. By expanding Eq.~(\ref{eq:CXY-74}) in the $T_{x_g}$, one obtains
\begin{align}
\int_{\tau}^1\frac{dz}{z^2}\int_{x_p}^1\frac{dx}{x}D_{h/g}\xi xG(x)\frac{\alpha_s}{2\pi}\int\frac{d^2x_\perp d^2y_\perp}{(2\pi)^2}(-2T_{x_g}(r_\perp))\left[{\mathbf H}_{2gg}^{(1)}+\int\frac{d^2b_\perp}{(2\pi)^2}{\mathbf H}_{2qg}^{(2)}\right]
\end{align}
with the hard coefficients
\begin{align}
{\mathbf H}_{2gg}^{(1)}=&\;{\cal H}_{2gg}^{(1)}-N_fT_Re^{-ik_\perp\cdot r_\perp}\delta(1-\xi)\int_0^1d\xi^\prime \left[\xi^{\prime2}+(1-\xi^\prime)^2\right]\left[\ln(1-\xi^\prime)^2+\ln\xi^{\prime2}\right]\non
&-N_c\delta(1-\xi)e^{-ik_\perp\cdot r_\perp}\int_0^1d\xi^\prime 2\left[\frac{\xi^\prime}{(1-\xi^\prime)_+}+\frac{1}{2}\xi^\prime(1-\xi^\prime)\right]\left[\ln\xi^{\prime2}+\ln(1-\xi^{\prime})^2\right],
\\
{\mathbf H}_{2gg}^{(2)}=&\;8\pi N_c\frac{[1-\xi(1-\xi)]^2}{\xi(1-\xi)_+}e^{-ik_\perp\cdot r_\perp}\Bigg[\frac{b_\perp\cdot r_\perp}{b_\perp^2r_\perp^2}e^{-ik_\perp\cdot b_\perp}\left(1+\frac{1}{\xi}e^{-i\left(\frac{1-\xi}{\xi}\right)k_\perp\cdot r_\perp}\right)\non
&-\frac{1}{\xi}\frac{(x_\perp-b_\perp)\cdot(y_\perp-b_\perp)}{(x_\perp-b_\perp)^2(y_\perp-b_\perp)^2}e^{-i\left(\frac{1-\xi}{\xi}\right)k_\perp\cdot(b_\perp-y_\perp)}\Bigg].
\end{align}
${\mathbf H}_{2gg}^{(1)}$ and ${\mathbf H}_{2gg}^{(2)}$ are equivalent to Eq.~(\ref{eq:Hgg1}) and Eq.~(\ref{eq:Hgg2}) respectively except for the power corrections as is the case in the $q\rightarrow q$ channel. As we show below, the $L_g$ term corresponds to the power corrections.

\subsection{The $q\rightarrow g$ channel}

Eq.~(82) in Ref.~\cite{Chirilli:2012jd} reads
\begin{align}
\frac{d\sigma_{(gq)}^{{\rm p}+A\rightarrow h/g+X}}{d^2p_{h\perp}dy}
=\sum_f\int_{\tau}^1\frac{dz}{z^2}\int_{x_p}^1\frac{dx}{x}D_{h/g}\xi xq_f(x)\frac{\alpha_s}{2\pi}\Bigg[
\int\frac{d^2x_\perp d^2y_\perp}{(2\pi)^2}S_{x_g}(x_\perp,y_\perp)\left[{\cal H}_{2gq}^{(1,1)}+S_{x_g}(y_\perp,x_\perp){\cal H}_{2gq}^{(1,2)}\right]\non
+\int\frac{d^2x_\perp d^2y_\perp d^2b_\perp}{(2\pi)^4}S_{x_g}^{(4)}(x_\perp,b_\perp,y_\perp){\cal H}_{4gq}^{(1)}\Bigg]
\label{eq:CXY-82}
\end{align}
where
\begin{align}
{\cal H}_{2gq}^{(1,1)}&=\frac{N_c}{2}\frac{1}{\xi^2}e^{-i\frac{k_\perp}{\xi}\cdot r_\perp}{\cal P}_{gq}(\xi)\ln\frac{c_0^2}{r_\perp^2\mu^2},\\
{\cal H}_{2gq}^{(1,2)}&=\frac{N_c}{2}e^{-ik_\perp\cdot r_\perp}{\cal P}_{gq}(\xi)\ln\frac{c_0^2}{r_\perp^2\mu^2},\\
{\cal H}_{4gq}^{(1)}&=-4\pi N_c{\cal W}\left(\frac{k_\perp}{\xi},k_\perp\right){\cal P}_{gq}(\xi)\frac{1}{\xi}\frac{x_\perp-y_\perp}{(x_\perp-y_\perp)^2}\cdot\frac{b_\perp-y_\perp}{(b_\perp-y_\perp)^2}
\end{align}
with ${\cal W}(k_{1\perp},k_{2\perp})=e^{-ik_{1\perp}\cdot(x_\perp-y_\perp)-ik_{2\perp}\cdot(y_\perp-b_\perp)}$. By expanding the quadrupole amplitude in the $T_{x_g}$, we cast Eq.~(\ref{eq:CXY-82}) into
\begin{align}
\sum_f\int_{\tau}^1\frac{dz}{z^2}\int_{x_p}^1\frac{dx}{x}D_{h/g}\xi xq_f(x)\frac{\alpha_s}{2\pi}\int\frac{d^2x_\perp d^2y_\perp}{(2\pi)^2}(-T_{x_g}(r_\perp))\left[{\mathbf H}_{2gq}^{(1)}+\int\frac{d^2b_\perp}{(2\pi)^2}{\mathbf H}_{2gq}^{(2)}\right]
\end{align}
with ${\mathbf H}_{2gq}^{(1)}={\cal H}_{2gq}^{(1,1)}+2{\cal H}_{2gq}^{(1,2)}$ and
\begin{align}
{\mathbf H}_{2gq}^{(1)}=-\frac{4\pi N_c}{\xi}{\cal P}_{gq}(\xi)\left[e^{-i\frac{k_\perp}{\xi}\cdot(b_\perp-y_\perp)-ik_\perp\cdot(x_\perp-b_\perp)}\frac{(b_\perp-x_\perp)\cdot(b_\perp-y_\perp)}{(b_\perp-x_\perp)^2(b_\perp-y_\perp)^2}
+e^{-i\frac{k_\perp}{\xi}\cdot(y_\perp-b_\perp)-ik_\perp\cdot r_\perp}\frac{(b_\perp-y_\perp)\cdot r_\perp}{(b_\perp-y_\perp)^2r_\perp^2}\right].
\end{align}
Eq.~(\ref{eq:Hgq1}) and (\ref{eq:Hgq2}) match ${\mathbf H}_{2gq}^{(1)}$ and ${\mathbf H}_{2gq}^{(2)}$ after taking large-$N_c$ limit, respectively. The subleading $N_c$ term remains in Eq.~(\ref{eq:Hgq2}) only gives a small contribution.

\subsection{The $g\rightarrow q$ channel}

Eq.~(87) in Ref.~\cite{Chirilli:2012jd} gives
\begin{align}
\frac{d\sigma_{(qg)}^{{\rm p}+A\rightarrow h/q+X}}{d^2p_{h\perp}dy}
=\int_{\tau}^1\frac{dz}{z^2}\int_{x_p}^1\frac{dx}{x}D_{h/q}\xi xG(x)\frac{\alpha_s}{2\pi}\Bigg[
\int\frac{d^2x_\perp d^2y_\perp}{(2\pi)^2}S_{x_g}(x_\perp,y_\perp)\left[{\cal H}_{2qg}^{(1,1)}+S_{x_g}(y_\perp,x_\perp){\cal H}_{2qg}^{(1,2)}\right]\non
+\int\frac{d^2x_\perp d^2y_\perp d^2b_\perp}{(2\pi)^4}S_{x_g}^{(4)}(x_\perp,b_\perp,y_\perp){\cal H}_{4qg}^{(1)}\Bigg]
\label{eq:CXY-87}
\end{align}
where
\begin{align}
{\cal H}_{2qg}^{(1,1)}&=\frac{1}{2}e^{-ik_\perp\cdot r_\perp}{\cal P}_{qg}(\xi)\left[\ln\frac{c_0^2}{r_\perp^2\mu^2}-1\right],\\
{\cal H}_{2qg}^{(1,2)}&=\frac{1}{2}\frac{1}{\xi^2}e^{-i\frac{k_\perp}{\xi}\cdot r_\perp}{\cal P}_{qg}(\xi)\left[\ln\frac{c_0^2}{r_\perp^2\mu^2}-1\right],\\
{\cal H}_{4qg}^{(1)}&=-4\pi {\cal W}\left(k_\perp,\frac{k_\perp}{\xi}\right){\cal P}_{qg}(\xi)\frac{1}{\xi}\frac{x_\perp-y_\perp}{(x_\perp-y_\perp)^2}\cdot\frac{b_\perp-y_\perp}{(b_\perp-y_\perp)^2}.
\end{align}
By linearizing Eq.~(\ref{eq:CXY-87}), one finds
\begin{align}
\int_{\tau}^1\frac{dz}{z^2}\int_{x_p}^1\frac{dx}{x}D_{h/q}\xi xG(x)\frac{\alpha_s}{2\pi}\int\frac{d^2x_\perp d^2y_\perp}{(2\pi)^2}(-T_{x_g}(r_\perp))\left[{\mathbf H}_{2qg}^{(1)}+\int\frac{d^2b_\perp}{(2\pi)^2}{\mathbf H}_{2qg}^{(2)}\right]
\end{align}
with ${\mathbf H}_{2qg}^{(1)}={\cal H}_{2qg}^{(1,1)}+2{\cal H}_{2qg}^{(1,2)}$ and
\begin{align}
{\mathbf H}_{2qg}^{(2)}=-&\frac{4\pi}{\xi}{\cal P}_{qg}(\xi)\left[e^{-i\frac{k_\perp}{\xi}\cdot(y_\perp-b_\perp)-ik_\perp\cdot(b_\perp-x_\perp)}\frac{(x_\perp-b_\perp)\cdot(y_\perp-b_\perp)}{(x_\perp-b_\perp)^2(y_\perp-b_\perp)^2}
+e^{-ik_\perp\cdot(y_\perp-b_\perp)-i\frac{k_\perp}{\xi}\cdot r_\perp}\frac{(b_\perp-y_\perp)\cdot r_\perp}{(b_\perp-y_\perp)^2r_\perp^2}\right].
\end{align}
${\mathbf H}_{2qg}^{(1)}$ and ${\mathbf H}_{2qg}^{(2)}$ are equivalent to Eq.~(\ref{eq:Hqg1}) and Eq.~(\ref{eq:Hqg2}) respectively in the large-$N_c$ limit.

\subsection{The $L_q$ and $L_g$ terms}

In Ref.~\cite{Watanabe:2015tja}, it has been shown that the so-called $L_q$ and $L_g$ terms play significant role in inclusive hadron production in p$A$ collisions at high $p_{h\perp}\gtrsim Q_s$. 
Regarding the $L_q$ term, Eq.~(10) in Ref.~\cite{Watanabe:2015tja} provides
\begin{align}
\frac{d\sigma_{L_q}}{d^2p_{h\perp}dy}=\sum_f\int_\tau^1\frac{dz}{z^2}D_{h/q}(z)x_pq_f(x_p)L_q(k_{\perp})
\end{align}
where 
\begin{align}
L_{q}(k_\perp)=\frac{\alpha_s N_c}{2\pi^2}\int\frac{d^2x_\perp d^2 y_\perp d^2b_\perp}{(2\pi)^2}e^{-ik_\perp\cdot (x_\perp-y_\perp)}\left[S_{x_g}(x_\perp,b_\perp)S_{x_g}(y_\perp,b_\perp)-S_{x_g}(x_\perp,y_\perp)\right]\non
\times\left[\frac{1}{u_\perp^2}\ln\frac{k_\perp^2u_\perp^2}{c_0^2}+\frac{1}{u_\perp^{\prime2}}\ln\frac{k_\perp^2u_\perp^{\prime2}}{c_0^2}-\frac{2u_\perp\cdot u_\perp^\prime}{u_\perp^2u_\perp^{\prime2}}\ln\frac{k_\perp^2|u_\perp||u_\perp^\prime|}{c_0^2}\right]
\label{eq:Lq}
\end{align}
with $u_\perp=x_\perp-b_\perp$ and $u_\perp^\prime=y_\perp-b_\perp$. By expanding the $L_q$ term in the $T_{x_g}$, Eq.~(\ref{eq:Lq}) can be cast into
\begin{align}
\frac{\alpha_s}{2\pi}\int\frac{d^2x_\perp d^2 y_\perp}{(2\pi)^2}e^{-ik_\perp\cdot r_\perp}(-T_{x_g}(x_\perp,y_\perp))\left[-N_c\left(\ln\frac{k_\perp^2r_\perp^2}{c_0^2}\right)^2-8N_c\pi\int\frac{d^2b_\perp}{(2\pi)^2}e^{-ik_\perp\cdot b_\perp}\frac{b_\perp\cdot r_\perp}{b_\perp^2 r_\perp^2}\ln\frac{k_\perp^2r_\perp^2}{c_0^2}\right].
\end{align}
This is completely equivalent to Eq.~(\ref{eq:doublelog-correction}).

Similarly, the $L_g$ term shown in Eq.~(23) in Ref.~\cite{Watanabe:2015tja} is given by 
\begin{align}
L_{g}(k_\perp)=\frac{\alpha_s N_c}{\pi^2}\int\frac{d^2x_\perp d^2 y_\perp d^2b_\perp}{(2\pi)^2}e^{-ik_\perp\cdot (x_\perp-y_\perp)}\left[-S_{x_g}(x_\perp,y_\perp)+S_{x_g}(x_\perp,b_\perp)S_{x_g}(b_\perp,y_\perp)\right]S_{x_g}(x_\perp,y_\perp)\non
\times\left[\frac{1}{u_\perp^2}\ln\frac{k_\perp^2u_\perp^2}{c_0^2}+\frac{1}{u_\perp^{\prime2}}\ln\frac{k_\perp^2u_\perp^{\prime2}}{c_0^2}-\frac{2u_\perp\cdot u_\perp^\prime}{u_\perp^2u_\perp^{\prime2}}\ln\frac{k_\perp^2|u_\perp||u_\perp^\prime|}{c_0^2}\right]
\end{align}
and eventually leads to 
\begin{align}
\frac{\alpha_s}{2\pi}\int\frac{d^2x_\perp d^2 y_\perp}{(2\pi)^2}e^{-ik_\perp\cdot r_\perp}(-2T_{x_g}(x_\perp,y_\perp))\left[-N_c\left(\ln\frac{k_\perp^2r_\perp^2}{c_0^2}\right)^2-8N_c\pi\int\frac{d^2b_\perp}{(2\pi)^2}e^{-ik_\perp\cdot b_\perp}\frac{b_\perp\cdot r_\perp}{b_\perp^2 r_\perp^2}\ln\frac{k_\perp^2r_\perp^2}{c_0^2}\right]
\end{align}
in the dilute limit. By replacing a factor of 2 in the front of the $T_{x_g}$ with $N_c/C_F$, the $L_g$ term agrees with Eq.~(\ref{eq:doublelog-correction-gg}).


\end{document}